# Tunable intervalence charge transfer in ruthenium Prussian blue analogue enables stable and efficient biocompatible artificial synapses

Donald A. Robinson, Michael E. Foster, Christopher H. Bennett, Austin Bhandarkar, Elizabeth R. Webster, Aleyna Celebi, Nisa Celebi, Elliot J. Fuller, Vitalie Stavila, Catalin D. Spataru, David S. Ashby, Matthew J. Marinella, Raga Krishnakumar, Mark D. Allendorf, and A. Alec Talin[*]

## Abstract

Emerging concepts for neuromorphic computing, bioelectronics, and brain-computer interfacing inspire new research avenues aimed at understanding the relationship between oxidation state and conductivity in unexplored materials. Here, we present ruthenium Prussian blue analogue (RuPBA), a mixed valence coordination compound with an open framework structure and ability to conduct both ionic and electronic charge, for flexible artificial synapses that reversibly switch conductance by more than four orders of magnitude based on electrochemically tunable oxidation state. Retention of programmed states is improved by nearly two orders of magnitude compared to the extensively studied organic polymers, thus reducing the frequency, complexity and energy costs associated with error correction schemes. We demonstrate dopamine detection using RuPBA synapses and biocompatibility with neuronal cells, evoking prospective application for brain-computer interfacing. By application of electron transfer theory to in-situ spectroscopic probing of intervalence charge transfer, we elucidate a switching mechanism whereby the degree of mixed valency between N-coordinated Ru sites controls the carrier concentration and mobility, as supported by DFT.

Mixed valence compounds (MVCs) that accommodate facile ion insertion and extraction have recently garnered intense interest for applications beyond energy storage such as neuromorphic computing, thermal and magnetic switching, and electrocatalysis.[1] This interest is motivated by the possibility to tune the electronic structure of the host via the coupled ion/electron insertion/extraction reactions characteristic of electrochemical charge transfer (CT). Among the various classes of MVCs, the continually growing family of Prussian blue analog (PBA) coordination frameworks[2,3,4] stand out due to their high (electro)chemical stability,[5,6,7] fast ionic transport,[8,9] biocompatibility and applications for biosensing,[10,11] process compatibility with electron and photolithography,[12,13] high electrocatalytic activity,[5,6] and synthetic tunability.[3,4] However, fundamental insight into the electronic conduction mechanism remains limited for most PBAs,[14,15] including the archetypal PB,[16,17,18,19] thus hindering their utilization for electronics applications. This is surprising, given the success of electron transfer theory in explaining the electronic properties of many mixed valence coordination compounds,[20,21,22] including cyanide-bridged complexes that mimic the molecular building blocks of PBAs.[23,24,25]

In this report, we describe a Ru-based Prussian blue analogue[26] (RuPBA) as both a model system for understanding redox-tunable electronic delocalization in extended mixed valence coordination compounds[27,28] and as an efficient active material for flexible electronic devices that rival organic electrochemical transistors (OECTs)[29,30] for emerging analog memory[31,32,33] and biointerfacing[34] applications. We elucidate the interrelationship between the electronic conductivity and the electrochemically tunable rate of intervalence CT using Marcus-Hush theory (MHT),[35,36] Boltzmann transport theory, and operando absorption spectroscopy, indicating that hole mobility >1 cm$^2$/(V·s) is achievable in this material. Application of MHT to the absorption spectra at different oxidation states reveals that electronic conduction predominantly occurs through a cross-pore (*CP*) charge transfer pathway between N-coordinated Ru sites, as depicted in Fig. 1a, consistent with DFT predictions. To show the potential of RuPBA in neuromorphic computing, we demonstrate electrochemical random-access memory (ECRAM) artificial synapses[31,37,38] that switch conductance over a range of >4 orders of magnitude with a large linear regime, low read/write currents, strong endurance, and excellent long-term state retention. The RuPBA ECRAM is inkjet printed on flexible substrates, biocompatible, and sensitive to dopamine. To our knowledge, the RuPBA represents the first practical alternative material to polymer blend based OECTs, offering promising avenues for innovation via its distinct ion/electron conduction mechanism and large synthetic flexibility of the PBA family of compounds.

Fig. 1a shows the idealized structure of RuPBA, depicted without $[Ru(CN)_6]^{4-}$ vacancies for simplicity, containing lithium ions intercalated within the square faces[39] of the cubic CN-bridged cages. The as-synthesized RuPBA contains 3 at.% of $K^+$ ions (Ru:K = 2.5:1), indicating that most of the Ru centers initially have a 2+ formal oxidation state. We expect the $K^+$ ions exchange with the $Li^+$ ions after the silica-based $Li^+$-containing ionogel electrolyte[40] layer is printed over the RuPBA layer. The transfer curve in Fig. 1c shows electrochemical tuning of the channel conductance from a lower limit of ~$9\times10^{-9}$ S to an upper limit of ~$2\times10^{-4}$ S. Based on the RuPBA film conductivity in the as deposited state ($4\times10^{-4}$ S/cm, Extended Data Fig. 1) and the measured conductance from Fig. 1c ($2\times10^{-5}$ S at $V_{gd}$ = 0 V), the electrochemically tunable conductivity range extends from ca. $10^{-6}$ to $10^{-2}$ S/cm. Interestingly, the lower conductivity at the fully reduced RuPBA (*fr*-RuPBA) state (~$10^{-6}$ S/cm) is close to that obtained by Xidis and Neff for *fr*-PB ($5\times10^{-7}$ S/cm).[18] We attribute the lower conductance limit at $V_{gd}$ ~ 0.6 V to that of *fr*-RuPBA, where all Ru centers have a 2+ formal oxidation state and the structure contains the maximum amount of lithium ions. Partial oxidation of *fr*-RuPBA causes $Ru^{II/III}$ mixed valency and conductivity to increase.

Most of the accumulated electrochemical gating charge should be proportional to the amount of $Ru^{II}$ species oxidized according to Faraday's law, and thus should also scale linearly with the charge carrier concentration. Analysis of the transfer data shows a voltage range where the conductance changes linearly with accumulated gate charge (see also Supplementary Information, Section S1). The trend suggests that only carrier concentration changes in the linear region, while mobility remains constant. Outside of this region, oxidation of the RuPBA channel either causes the mobility to increase (0.6 V ≲ $V_{gd}$ ≲ -0.1 V) or decrease (-0.3 V ≲ $V_{gd}$ ≲ -0.6 V). Upon oxidizing the channel further, the conductance decreases, indicating that a mixed valence form is the most conductive (Extended Data Fig. 2).

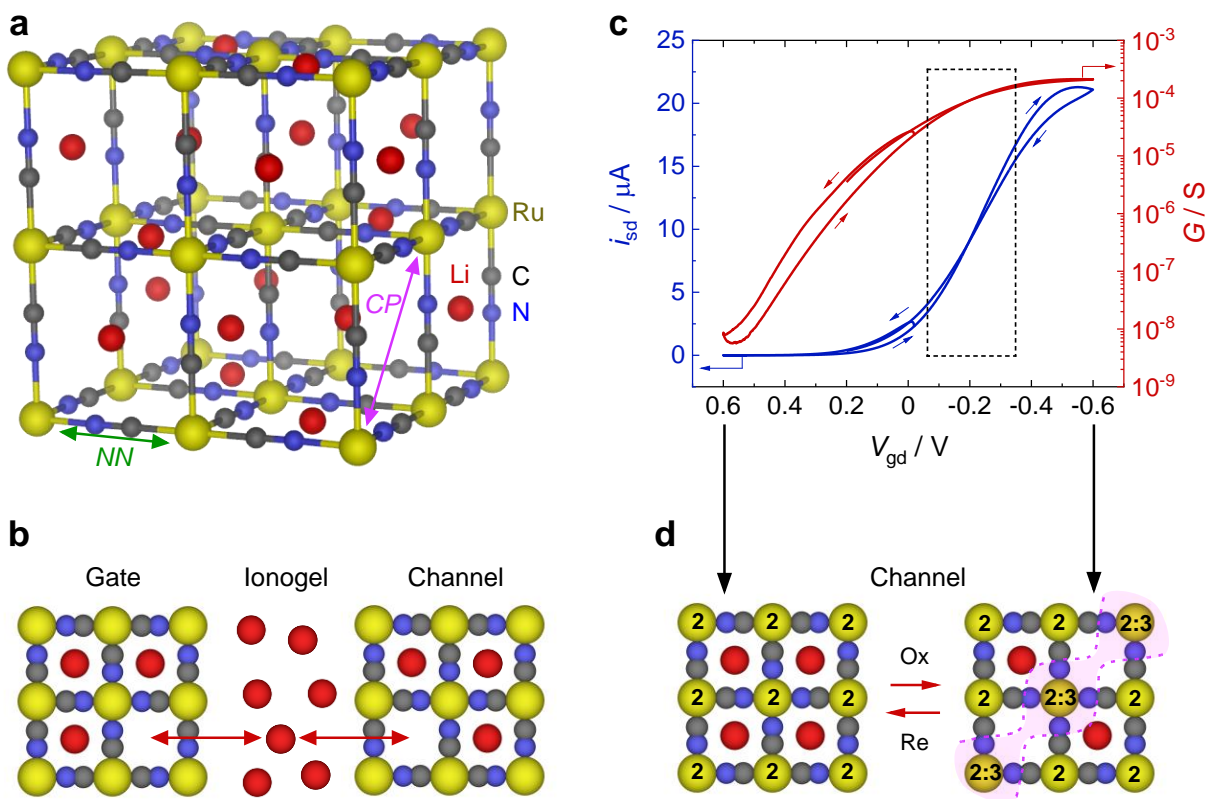

**Fig. 1.** Electrochemical doping of RuPBA. **a**, Idealized vacancy-free structure of partially lithiated RuPBA, $Li_3Ru_2[Ru(CN)_6]_2$, showing proposed cross-pore (*CP*) and nearest neighbor (*NN*) charge transfer paths. **b**, Schematic representing reversible $Li^+$ ion insertion into the RuPBA film at the gate and channel separated by a $Li^+$-containing ionogel electrolyte (See Supplementary Information, section S2, for a detailed description of the ECRAM device architecture). **c**, Transfer curve for a RuPBA/Li-ionogel device showing the resulting source-drain current of the channel ($i_{sd}$) on a linear scale (left y-axis, blue) and channel conductance on a logarithmic scale ($G$, right y-axis, red) when scanning the gate-drain voltage ($V_{gd}$) at 30 µV/s and applying a source-drain bias of 0.1 V. The dashed box designates the region where $i_{sd}$ increases linearly as a function of electrochemical charge. **d**, Scheme conveying the change in Ru valency and degree of lithiation as the channel undergoes oxidation (Ox) and reduction (Re). Numeric labels on Ru atoms represent formal oxidation states and valence mixing. The region shown in pink highlights the proposed crosspore mixed valence conduction pathway, where electron delocalization occurs between neighboring N-coordinated Ru centers while C-coordinated Ru centers maintain localized 2+ formal oxidation states.

## Evaluation of ECRAM device performance

Consistent with the transfer characteristics (Fig. 1c), the voltage-controlled pulsed programming of an ECRAM device with a 0.75 ms 'write' pulse duration shows that the potentiation ($G\uparrow$) and depression ($G\downarrow$) ramps (Fig. 2a) are highly linear, which is desirable for high accuracy during training.[32,33] Devices fabricated with a proton-conducting polymer-based ionogel electrolyte show better switching behavior in terms of linearity and symmetry, but switch more slowly, requiring more than 10-fold longer write pulse durations and twice as many pulses to cycle over a range of $G_{max} = 2G_{min}$ (Extended Data Fig. 3). The conductance range of 80-160 nS in Fig. 2a corresponds to a read current range of 8-16 nA at the applied read voltage ($V_r = 0.1$ V). The combination of low current and near-linear conductance tuning is ideal for energy-efficient programming.[32]

State retention is essential for nonvolatile synaptic memory. As shown in Fig. 2b, the programmed conductance over the entire range, 0.65-2.0 µS ($G_{max}/G_{min} \sim 3$), has a retention time of ~600 s (defined as $dG < 1\%$). The loss in retention at $G > G_0$ in Fig. 2b indicates that the more oxidized RuPBA states are less stable than the reduced states. However, the conductance remains stable within ±1% over a duration of ~$10^4$ s for the range below the initial state, 0.65-1.5 µS ($G_{max}/G_{min} \sim 2$), an improvement in comparison to the ~250 s retention time reported for PEDOT:PSS synapses programmed within a similar $G_{max}/G_{min}$ range under inert environment.[41] PEDOT:PSS channels contain polyethyleneimine (PEI), which acts as an electron donor to annihilate electrochemically generated holes until the conductivity decays to its initial state. When charged to a lower conductivity, the PEI additive diffuses out of the channel and into the electrolyte, causing conductivity to increase over time towards the initial state.[41,42] Given standard neural network training, the negative impacts of such volatility on network accuracy and energy efficiency might be considerable, *e.g.*, a deep neural network based on filamentary ReRAM where a ~5% conductance change over a period of $10^4$ s at elevated temperature led to ~16 % loss in network accuracy, and could only be mitigated through a correction scheme applied every ~$10^3$ s.[43,44] Yet, the conductance of RuPBA, when programmed at the lower range ($G < G_0$), does not trend back towards $G_0$ because RuPBA does not rely on a redox buffer such as PEI. This implies that, with only mild loss of analog range, non-volatile programming in the range of 0.65-1.5 µS can still enable efficient *in-situ* learning and retention for embedded biocompatible applications, with only minor alterations to normal training needed (*e.g.*, an array calibration step after each epoch to detect outlier weights exceeding some threshold above $G_0$, where they can be clipped/RESET). As most neural network weights are low significance (low-conductance), the hardware mapping for this scheme is natural and could lead to low overall error profiles at inference stage.[45] Meanwhile, the range clipping would not be a major detriment to learning performance, since a compressed range of 6-8 bits writable space is more than sufficient for most online learning applications using emerging non-volatile memory devices[46,47] even when considering write noise in the loop.[48]

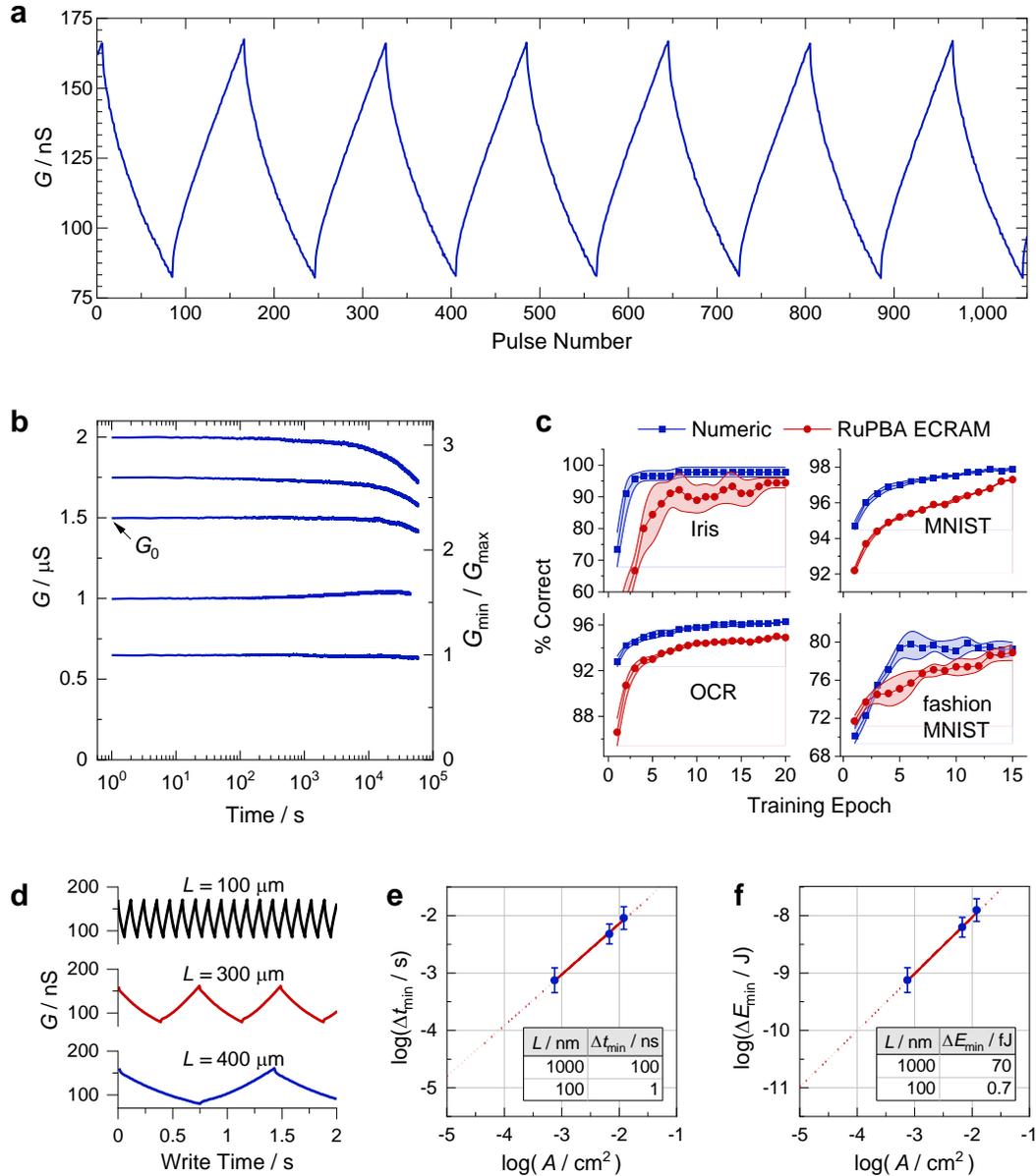

**Fig. 2.** Li-RuPBA ECRAM performance. **a**, Channel conductance during pulsed programming at $V_{gd} = +1$ V/-1 V for depression/potentiation at 0.75 ms gate pulse duration. **b**, Long term retention of 5 programmed conductance states. **c**, Simulated performance for neuromorphic tasks. Error bands represent standard deviation. **d**, Pulsed conductance programming with different channel dimensions, as labelled. 'Write Time' represents the total time that voltage is applied to the gate. Log-log plots showing relationship of channel area, $A$, to **e**, switching times, $\Delta t_{min}$, and **f**, switching energies, $\Delta E_{min}$, required to change the channel conductance by an average value of ±1.0 nS. Inset tables

show values obtained by extrapolation of the linear fits (red lines). The durations of the applied write pulses were 0.75 ms, 6 ms, and 10 ms in order of increasing area; this was done to achieve roughly equal numbers of conductance states per ramp, ~70 states for each of the three devices. The channel length:width (L:W) aspect ratio of devices for parts **d-f** is 1:7.5, keeping the RuPBA film thickness constant at ~0.6 μm. Error bars represent the standard deviations for parts **e-f**. See Table S3 of the Supplementary Information for number of switching pulses analyzed per device. All measurements were performed under $N_2$ atmosphere at T = 300 K.

The channel conductance can be repeatedly cycled using $10^6$ switching pulses without altering switching characteristics (Extended Data Fig. 4). From this data, we evaluate online learning by simulating a hardware multi-layer perceptron populated with RuBPA synapses. Due to the existence of a smooth, linear workable region optimal for on-chip backpropagation,[49] the RuBPA synapse generally performs excellently in the context of online learning. As shown in Fig 2c, on smaller tasks, such as the Iris flower classification problem[50] and the optical character recognition (OCR) task,[51,52] RuPBA synapses achieve close agreement with ideal synapses (numerical). On larger neuromorphic tasks such as the fashion-MNIST[53] and MNIST task,[54] the gap between ideal and RuPBA devices diverges slightly more. Overall, our results are comparable to those of other leading linear and symmetric nonvolatile memory candidates (Supplementary Information, Section S3).[31,33,37,55,56,57]

The programming speed and energy efficiency of an ECRAM device depends on the electrochemical current of the 'write' step when a gate voltage is applied to change the redox state of the channel. Write currents should be minimized to reduce the energy cost.[31] However, higher faradaic write current densities should increase the ECRAM switching rate (faster charging). Fig. 2d shows conductance vs. write time traces for three devices with different channel areas, $A$, but a similar length:width aspect ratio ($L:W$ = 1:7.5). The smallest device clearly operates more quickly, completing 12 potentiation/depression cycles before the larger device can finish the first cycle. The average times and energies required to switch the conductance by a magnitude of 1 nS are plotted versus channel area in Fig. 5e-f (See Supplementary Information, section S4). The smallest device achieves a switching time of 0.75 ms, comparable to the duration of action potentials in biological neurons.[58] The relationships are highly linear with log-log slopes of 1. From extrapolation, we predict that RuPBA ECRAM devices of similar architecture with 100 nm channel length should switch on the timescale of ~1 ns, much faster than biosynapses, with ~1 fJ write energy, a lower energy cost compared to ~10 fJ per synaptic event in biological synapses.[58] We anticipate that RuPBA devices fabricated with thinly layered channel/electrolyte/gate vertical architecture[37] would significantly lower the electrolytic gating resistance and increase the overall switching efficiency.[32]

**Potential for biosensing and brain-computer interfacing**

RuPBA undergoes reduction upon reaction with dopamine (DA), an important neurotransmitter. Fig. 3a shows the design for a modified ECRAM device that allows DA to react with the RuPBA gate while preventing direct reaction with the channel. Fig. 3d shows that the device conductance decreases by $\Delta G$ when incrementally increasing the DA concentration, leading to the dose-response calibration curve. (see Supplementary Information, section S5, for a more detailed description of sensor operation). The RuPBA synapse achieves a 4 µM limit of detection, comparable to PEDOT:PSS-based OECTs.[34, 59]

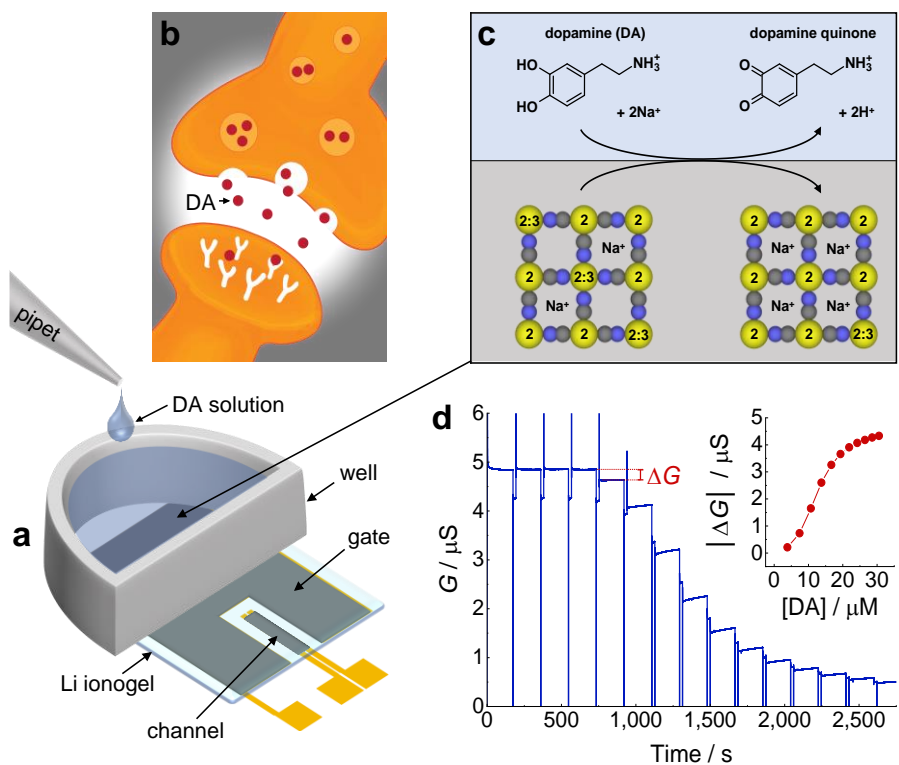

**Fig. 3.** Dopamine (DA) sensing with modified ECRAM cell. **a**, Sketch of modified device for dopamine sensing. **b**, Illustration of dopamine release from vesicles of a presynaptic neuron (top) to receptors of a postsynaptic neuron (bottom). **c**, Schematic of dopamine reaction with RuPBA in the portion of the device gate enclosed by the well. **d**, Conductance trace during incremental additions of dopamine and inset showing corresponding dose-response plot.

In the inset, $|\Delta G| = |G - G_0|$, where $G_0$ (= 4.85 μS) is indicated by the topmost dashed horizontal red line on the conductance trace.

The sensitivity of RuPBA to dopamine establishes a basis to develop hybrid bio/artificial synapses.[34] For successful bio-interfacing, neuronal cells should be able to grow and thrive on RuPBA films. We find that cells seeded onto RuPBA films divide at a rate comparable to control cells, and that their viability is as high as control cells (Extended Data Fig. 5a). Optical and scanning electron micrographs of the cells show interaction with and adherence to the RuPBA substrate (Extended Data Fig. 5b-e). We conclude that RuPBA is nontoxic and nonrefractory to cell growth. The compatibility of RuPBA with live neuronal cells signifies a promising first step towards utilizing electrochemical transistors to integrate biological and artificial analog neural networks.

**Spectroscopic probing of RuPBA electronic structure during switching**

In-situ UV/Vis/NIR spectroscopy provides a means to probe the electronic structure of RuPBA during electrochemical cycling. The visible color change of an electrochromic RuPBA film on FTO/glass is shown in the Supplementary Information, Video S1. The film changes color from beige at the more reduced state to blue grey at the more oxidized state. Fig. 4a shows the evolution of absorption spectra as a RuPBA/ITO electrode is scanned from a fully reduced state at -0.2 V to a partially oxidized state at 1.0 V (versus Ag/AgCl wire). Following Behera et al., we assign the bands at ca. 30,000 cm$^{-1}$ and 6000 cm$^{-1}$ to the metal-to-ligand charge transfer (MLCT) and intervalence charge transfer (IVCT) reactions, respectively.[26] As the film is oxidized, the MLCT intensity decreases while the IVCT intensity increases. Like the conductance vs gate voltage trend of Fig. 1c, the relationship of IVCT absorbance to electrode potential in Fig. 4b is sigmoidal and reversible. The absorbance vs potential response is consistent over multiple cycles in this 1.2 V potential window. Fig. 4c shows the interrelationship of applied gate voltage, IVCT absorbance, and conductance when switching conductance across four orders of magnitude in a RuPBA ECRAM device. The correlation of conductance to IVCT absorbance signifies that the electronic conductivity of RuPBA is inextricably linked to the rate of intervalence charge transfer.

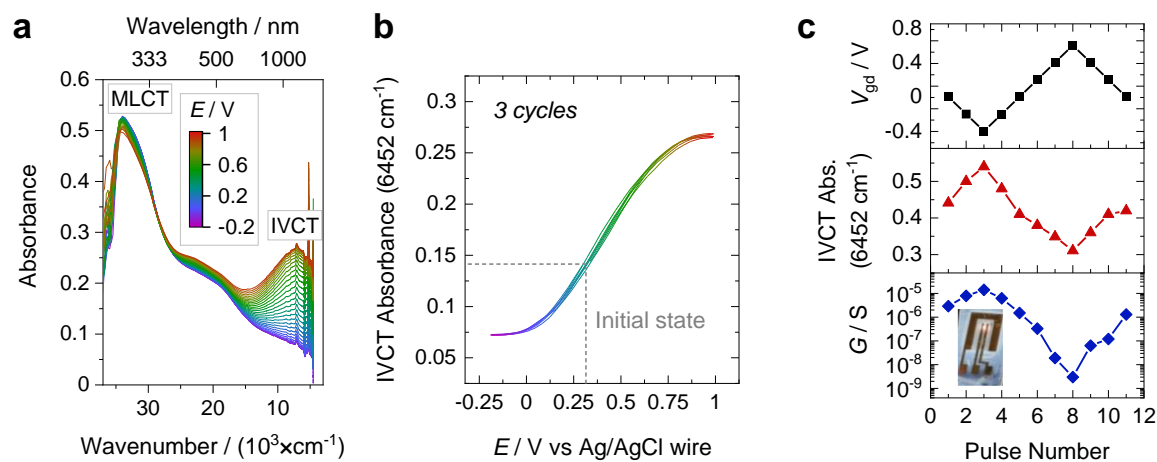

**Fig. 4.** In-situ UV-Vis-NIR absorption spectroscopy of RuPBA films during voltammetric cycling. **a-b**, Results from cyclic voltammetry of RuPBA film on indium tin oxide (ITO)-coated glass in contact with liquid electrolyte containing 1.0 M LiFSI dissolved in Pyr$_{14}$TFSI, scan rate of 50 µV/s. **a**, Overlaid spectra recorded as the potential, $E$, was scanned from -0.2 to 1.0 V (vs Ag/AgCl wire quasireference electrode). **b**, Absorbance monitored at 6452 cm$^{-1}$, as taken from the 2$^{nd}$ scan cycle, with the initially recorded open circuit potential at equilibrium, 'initial state', marked by the gray dashed lines. **c**, Measurements of the conductance and IVCT absorbance in the channel region of an ECRAM device after potentiostatic charging at varied gate voltages ($V_{gd}$). Conductance is shown on a logarithmic scale. Inset image on the conductance plot shows the optical path from the tungsten halogen light source through the device channel.

When extending the applied potential to 1.4 V, we observe an additional trend in the IVCT region of the UV/Vis/NIR spectra (Extended Data Fig. 6d). The lowest energy maximum at ca. 6450 cm$^{-1}$ follows the trend as discussed above from -0.2 to 1.0 V, but then decreases in absorbance from 1.0 to 1.4 V. The absorbance at ca. 13,000 cm$^{-1}$, however, continues to increase as RuPBA undergoes oxidation to 1.4 V. The two observed trends reveal that two distinct IVCT processes occur in RuPBA, and that the absorbance of the lower energy IVCT correlates with conductivity.

**Mechanistic interpretation**

The strong correlation of IVCT absorbance with conductivity supports a mechanism whereby electronic transport in RuPBA occurs through adiabatic intramolecular CT reactions. As shown in Fig. 5a, deconvolution of the UV/Vis/NIR spectrum for a film of the as-synthesized RuPBA (*as*-RuPBA) suggests at least seven different overlapping optical absorption bands in the wavenumber ($v$) range of 35,000 to 4,000 cm$^{-1}$, consistent with the previous report.[26] We assign the two lowest energy bands at ca. 8400 cm$^{-1}$ (1.0 eV) and 5100 cm$^{-1}$ (0.6 eV) to two different IVCT transitions, as depicted in Fig. 5a. The higher energy band (IVCT$_{NN}$) corresponds to electron transfer from N-coordinated Ru (N-Ru) to its nearest neighbor, C-coordinated Ru (C-Ru), as separated by the cyanide bridge. The lower energy band (IVCT$_{CP}$) corresponds to 'cross-pore' transfer between two N-coordinated Ru$^{II}$/Ru$^{III}$ redox pairs. Our interpretation of these two separate IVCT bands agrees with that of isolated trinuclear and tetranuclear CN-bridged Ru complexes.[23,60,61,62] Both IVCT bands are best fit using an asymmetric peak shape, which indicates that the *as*-RuPBA fits in a transition regime of electronic delocalization according to Robin-Day classification,[63] exhibiting properties typical for both class 2 (semiconductor) and class 3 (metallic) systems, or the so-called "class 2B" system.[22] In contrast, PB belongs to the class 2A regime involving more localized charges.[63]

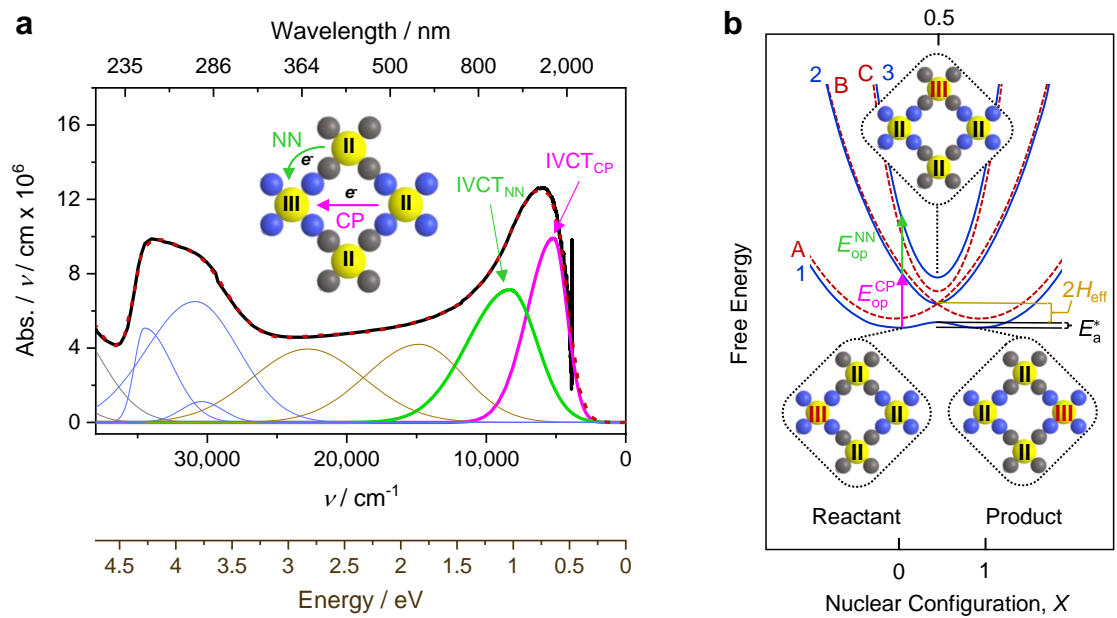

**Fig. 5.** Interpretation of optical charge transfer bands from UV/Vis/NIR absorption spectrum of RuPBA. **a**, Reduced absorption spectrum and fitted asymmetric gaussian peaks, highlighting two different IVCT bands for transfer between nearest neighbor cyanide bridged Ru centers (IVCT$_{NN}$) and cross-pore between N-coordinated Ru centers (IVCT$_{CP}$), as depicted in the inset scheme. The absorbance is divided by $\nu$ to accurately determine the reorganization energy and coupling in accord with theory.[22] **b**, Schematic representation for potential energy surface diagram of IVCT based on a three-state model. Red dashed lines labeled A, B, and C convey three parabolic overlapping diabatic energy surfaces, where the energy minima for A,B are equivalent and occur at $X = 0,1$. The higher energy minimum for C occurs at $X = 0.5$. Blue lines labeled 1, 2, and 3 represent the adiabatic states resulting from electronic coupling between N-coordinated donors, acceptors, and hexacyanoruthenate bridge states. The IVCT$_{CP}$ product and reactant states are of equivalent ground state energy due to symmetry. The third possible state (state 3), formed from IVCT$_{NN}$, is thermodynamically disfavored. $E_{op}^{NN}$ and $E_{op}^{CP}$ represent $hv$ for the deconvoluted peak maxima of IVCT$_{NN}$ and IVCT$_{CP}$, respectively.

The effective donor-acceptor coupling, $H_{eff}$, and activation energy, $E_a^*$, were determined based on the IVCT$_{CP}$ peak energy and shape following an equivalent two-state model[22] for the three-state system depicted in Fig. 5b (see Supplementary Information, Section S6). The fit of the IVCT$_{CP}$ band for *as*-RuPBA gives $E_{op}^{CP} = \lambda = 0.64$ eV and a large coupling, $H_{eff} = 0.22$ eV; this leads to $E_a^* = 16$ meV, which is less than $k_BT$ at room temperature. The IVCT$_{CP}$ rate constant, $k_{ct}$, is then

$$k_{ct} = \kappa v_n \exp(-E_a^*/k_BT) \qquad (1)$$

where $\kappa$ is the probability of electron transfer at the configuration coordinates ($X$) when the activation barrier is minimized and $v_n$ is a nuclear vibration frequency,[21] here assigned as the cyanide bridging ligand vibration (ca. 2130 cm$^{-1}$) shown in Extended Data Fig. 7, giving $v_n = 6.4 \times 10^{13}$ s$^{-1}$. The calculated electronic frequency[20,64] ($v_{el} = 1.0 \times 10^{15}$ s$^{-1}$, see Supplementary Information, section S6) is much higher than $v_n$, such that $\kappa$ is essentially unity, as expected. This leads to a very fast rate constant of $k_{ct} \approx 3 \times 10^{13}$ s$^{-1}$, only slightly lower than $v_n$, further supporting a Robin-Day class 2B assignment for *as*-RuPBA, but very close to class 3 delocalization.[22] The $k_{ct}$ value corresponds to an IVCT timescale of 30 fs, which is within the sub-picosecond range measured for CT in PBAs by time-resolved spectroscopy.[65,66,67]

Rosseinsky and coworkers presented a simple expression for IVCT-based conductivity of PB following the treatments described by Austin-Mott and Robin-Day.[63,68,69] We propose a similar relationship below.

$$\sigma = pe^2d^2k_{ct}/2k_BT = ep\mu_h \qquad (2)$$

Here, the conductivity ($\sigma$) depends on the number density of hole charge carriers ($p$) and the distance between reactant centers ($d$). For IVCT$_{CP}$, the shortest distance between N-Ru centers is $d \approx 7.4$ Å based on the XRD pattern of RuPBA (Extended Data Fig. 8a). Eq. 2 also shows how $k_{ct}$ and $d$ relate to the hole mobility; $\mu_h = ed^2k_{ct}/2k_BT$. For *as*-RuPBA, $\mu_h \approx 4$ cm$^2$/(V·s), which is comparable to that reported for PEDOT:PSS.[70] Because both IVCT$_{CP}$ absorbance and conductivity increase as the RuPBA partially oxidizes, we propose that $p$ must relate to the concentration of mixed valence N-Ru sites, as depicted in Fig. 5. The maximum carrier concentration occurs when the valence mixing is 1:1 II:III between N-Ru centers, corresponding to a net mixing of 3:1 II:III for all Ru centers, assuming an ideal defect-free lattice.

Closer analysis of the spectra in Fig. 4a reveals that the IVCT band manifold not only increases in absorbance, but also shifts by more than 0.3 eV to lower energy throughout the course of oxidation from $fr$-RuPBA to the most conductive mixed valence state at 1.0 V, signifying that $\lambda$ and $E_a^*$ decrease while $k_{ct}$ and $\mu_h$ increase (see Supplementary Information, Section S6). Spectroscopically determined parameters at varied $E$ (Extended Data Table 1) indicate that mobility increases from 1 to 4 cm$^2$/(V·s) when oxidizing RuPBA over a small potential range from $E$ = 0 to 0.1 V. However, the mobility does not vary significantly at potentials between 0.1 V and the most conductive state ($E$ = 1.0 V), implying that the increasing carrier concentration from oxidation is the sole contributor to conductance switching in the $V_{gd}$ range roughly between 0 and -0.5 V (Fig. 1c). The trend of increasing, constant, then decreasing mobility during partial oxidation of $fr$-RuPBA agrees well with the interpretation of the transfer curve above (Fig. 1c and Supplementary Information, Fig. S1).

The partial density of states (pDOS) computed by DFT for different oxidation/lithiation states of RuPBA is shown in Fig. 6. The 'fully reduced,' 'partially reduced', and 'neutral' forms correspond to compositions of Li$_x$Ru$_2$[Ru(CN)$_6$]$_2$, where $x$ = 4, 3, and 2, respectively. The partially reduced form ($pr$-RuPBA), as depicted in Fig. 1a, is predicted to have a direct band gap of 0.50 eV, matching the experimentally determined IVCT$_{CP}$ maximum for the most conductive state (see Extended Data Table 1). Such agreement between computed band gap and measured IVCT excitation energy was also reported for PB.[71] The fully reduced form is predicted to have a large band gap (4 eV), comparable to the energy of the observed MLCT band ($h\nu_{max}$ ~ 4 eV). The calculated hole effective masses ($m_h$) for RuPBA range from 0.4 to 0.7 (see Extended Data Table 2); these values are similar to known semiconductors such as p-type germanium.[72] The smallest $m_h$ value corresponds the $pr$-RuPBA state. Based on scattering theory (see DFT Computation for details), we predict a hole charge mobility for $pr$-RuPBA of 5 cm$^2$/(V·s), which is in remarkable agreement with our experimental estimate of 4 cm$^2$/(V·s) from the MHT analysis above. The pDOS for $pr$-RuPBA reveals a delocalized electronic structure, but valence and conduction band states closest to the Fermi energy remain primarily associated with N-Ru sites. This suggests that the lowest energy excitation is between N-Ru sites, which agrees with our interpretation of the IVCT spectra (Fig. 5). Thus, both band and electron transfer theories support a mechanism whereby the electrochemically tunable degree of mixed valency between N-Ru centers directly relates to charge carrier concentration.

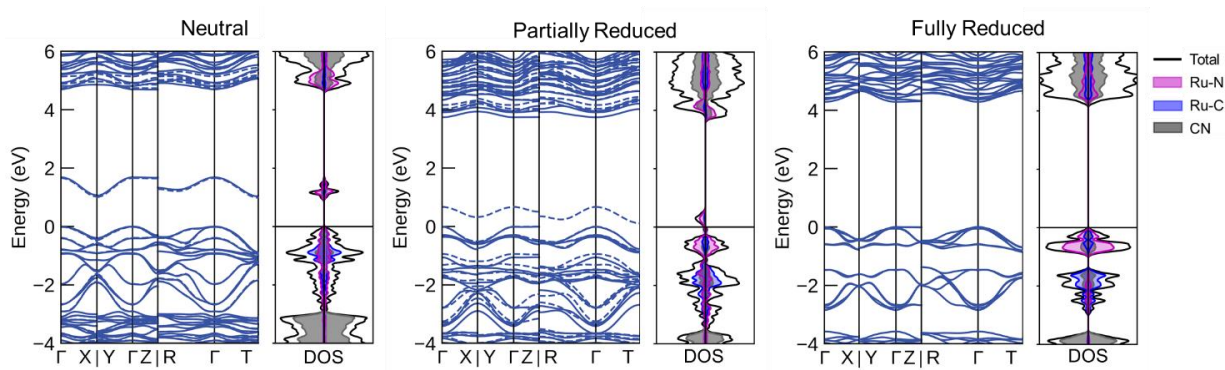

**Fig. 6.** DFT/HSE06 band structures and density of states of RuPBA in the neutral, partially reduced and fully reduced states. The hole effective masses are reported in Extended Data Table 2.

We also compute pDOS and $m_h$ values for PB at different redox states for comparison (Supplementary Information, Section S7). The $m_h$ values for PB (Table S4) are larger than those of RuPBA, ranging from 1.3 to 2.9, in agreement with previously published values,[73] which is consistent with the predicted flat bands and localized states (Fig. S8), and in agreement with measurements by Long that showed RuPBA is substantially more conductive than PB.[26] The differences between PB and RuPBA band structures primarily arise from their different electron spin configurations (Fig. S9). Unlike the high/low spin configuration of PB,[19] both N-coordinated and C-coordinated metal atoms of RuPBA are low spin, allowing for better energy alignment between neighboring Ru atoms and a more delocalized electronic structure (Fig. S10-11).

**Effect of grain boundaries**

The estimated mobilities presented above specifically refer to electron-hole transport within the bulk CN-bridged framework. Assuming a hole density ($p$) of ca. $10^{21}$ cm$^{-3}$, we estimate a bulk crystal conductivity greater than 100 S cm$^{-1}$ for the most conductive state (*pr*-RuPBA), which is ~10,000 times greater than the measured maximum conductivity. The inkjet-printed films, however, are composed of RuPBA nanocrystallites (Extended Data Fig. 8b). Measurements of DC conductivity as a function of temperature for *as*-RuPBA give an activation energy of 0.18 eV (Extended Data Fig. 1), which is much higher than the activation energies estimated above from spectroscopy. CT across grain boundaries likely plays a limiting role in the observed film conductivity.[74] Importantly, this activation energy is roughly ¼ of the measured reorganization energy for the IVCT$_{CP}$ band ($\lambda = 0.64$ eV), consistent with very weak coupling and polaron hopping[69,75] between charge-localized N-Ru centers at the interfaces between neighboring nanocrystals. We propose that electronic coupling is much weaker across grain boundaries due to the absence of NCRuCN bridges between N-coordinated Ru centers.

**Outlook**

We have shown that the class of active materials for electrochemical transistors extends well beyond extensively studied organic polymers and transition metal chalcogenides into the realm of extended coordination frameworks like RuPBA. Our findings open a window to study electrochemically tunable conductivity in related mixed valence coordination structures such as other PBAs, coordination polymers, and metal-organic frameworks. Two different theoretical approaches, electron transfer theory and Boltzmann transport theory, independently predict similar carrier mobilities within the RuPBA bulk. Our spectroscopic analysis emphasizes that the Marcus-Hush model is not only applicable to CT kinetics in small molecules, but also provides physical insight into mixed valence conductors that belong to the Robin-Day class 2-3 transition regime,[22,63] exhibiting characteristics of both intervalence hopping and band-like transport due to an intermediate degree of electron delocalization. Both theoretical treatments also support the proposed IVCT$_{CP}$ transport pathway. We anticipate major improvements to ECRAM and other electronic applications as devices are made smaller and look forward to taking advantage of RuPBA's sensitivity to neurotransmitters to achieve communication between biological and artificial neurons.

## Methods

All reagents and solvents were purchased from Sigma-Aldrich unless otherwise noted. All water was filtered and deionized to 18.2 MΩ·cm resistivity before use (*PureLab flex* filtration system, ELGA LabWater). Custom-designed shadow masks for e-beam deposition were purchased from Stencils Unlimited, LLC.

**Synthesis of RuPBA**

RuPBA was synthesized following the hydrothermal method reported by Long and coworkers.[26] Ruthenium(III) acetylacetonate, $Ru(C_5H_7O_2)_3$, (0.500 g, 1.24 mmol) was added to a 20 mL aqueous solution containing 0.260 g (0.310 mmol) of potassium hexacyanoruthenate(II), $K_4[Ru(CN)_6]$. The mixture was stirred for 1 hr at room temperature and subsequently transferred to a 45-mL PTFE-lined acid digestion vessel (Parr, model 4744) and heated at 175 °C for 96 hr. The resulting black precipitate was isolated by centrifugation and washed three times with 70 mL of water. To remove $Ru(C_5H_7O_2)_3$ starting material and potential organic byproducts, the precipitate was rinsed once with 70 mL acetone, three times with 40 mL toluene, and three additional times with 40 mL acetone, using a sonication bath to redisperse the sample in each rinsing solvent and a centrifugation rate of 6000 rpm to collect the precipitate. After rinsing 5 times with 40 mL aliquots of water by sonication and centrifugation (8700 rpm), the final precipitate was redispersed in water and separated into two equal portions. The portion for XRD and elemental analysis was isolated by centrifuge and dried at 60 °C for two days before characterization. Found composition (weight %): K, 6.9; Ru, 45.7; C, 17.6; N, 15.1; H, 1.9. The second portion was isolated by centrifuge and redispersed in DMF to give a 10 mg/mL suspension for preparation of the RuPBA ink.

**Preparation of RuPBA ink**

Cyclohexanol (2.5 mL) was added to a 5 mL aliquot of 10 mg/mL RuPBA suspension in DMF and sonicated for 10 min. The resulting suspension was centrifuged at 2500 rpm for 5 min to collect and discard large RuPBA aggregates. The remaining suspension, ~4 mg/mL RuPBA in 2:1 DMF:cyclohexanol (v:v), was used as the RuPBA ink.

**Preparation of Li ionogel ink**

Lithium bis(fluorosulfonyl)imide, LiFSI, (2.035 g) was dissolved in 10 mL of 1-butyl-1-methylpyrrolidinium bis(trifluoromethanesulfonyl)imide (Pry$_{14}$TFSI) ionic liquid by stirring overnight in an Ar-filled glovebox to give a 1.0 M LiFSI in Pry$_{14}$TFSI. Preparation of the sol-gel precursor for the ionogel ink was based on the method described by Dunn and coworkers.[40] A 300 μL volume of tetraethylorthosilicate (TEOS) was added to 300 μL of triethoxyvinylsilane (TEVS) and sonicated for 15 min. While under sonication, 300 uL of formic acid was added to the TEOS/TEVS mixture and allowed to sonicate for 15 min to give the sol-gel precursor. A 550 μL aliquot of the sol-gel precursor was then added to 450 uL of the ionic liquid (1.0 M LiFSI in Pry$_{14}$TFSI) and sonicated 15 min to give a clear solution, which was subsequently diluted with 1 mL ethanol for use as the Li ionogel ink.

**Preparation of protic ionogel ink**

A solution containing 17.6 wt% protic ionic liquid (diethylmethylammonium trifluoromethanesulfonate, DEMA:TfO, 98%, IOLITEC), 4.4 wt% poly(vinylidene fluoride-*co*-hexafluoropropylene), and 78 wt% dimethylacetamide (DMA) was prepared by sonication and subsequently diluted by a factor of 4 with additional DMA solvent.

**ECRAM device fabrication**

Gold electrode patterns for ECRAM devices and Van der Pauw samples were fabricated on nanosilica-coated polyethylene terephthalate (PET) printing media (Novele™, Novacentrix) via e-beam evaporation using custom-designed shadow masks (Stencils Unlimited) by depositing 5 nm Ti adhesion layer, followed by 150 nm Au top layer. RuPBA and ionogel layers were printed over the Au-patterned PET substrates using a Dimatix Materials Printer (DMP-2850) with DMC-11610 ink cartridges. Removable Scotch tape was used to adhere the backside of the PET substrate to an SiO$_2$-coated Si wafer. The fiducial camera of the DMP was used to precisely align the RuPBA prints to the Au electrode pattern array.

RuPBA films for ECRAM devices were printed using a cartridge temperature of 29 °C and substrate temperature of 55 °C. Device channels were printed with one jet nozzle at 20 μm drop spacing for optimal film quality and feature resolution, with a total of 10 print cycles. Channel films included ~75 μm of overlap with the Au contacts. For the device gates, 15 RuPBA print cycles were performed using 11 nozzles at a drop spacing of 31 μm. Two-minute rest periods were inserted in between prints for each device row to keep the cartridge temperature from exceeding 32 °C, which was found to help prevent nozzle clogging. The resulting RuPBA films were then dried at 60 °C for at least 48 hours. After realignment of the substrate, ionogel films were printed at room temperature with 20 μm drop spacing using 3 jetting nozzles and at least 6 print layers to ensure full coverage.

ECRAM devices for dopamine sensors included an additional 4 by 4 mm area to the gate to serve as the base of the reservoir for the dopamine solution. For these devices, RuPBA was printed over the entire area of Au gate electrode and the channel region, similar to the other ECRAM devices. The ionogel was printed over the region shown in Fig 3a, leaving a 4 by 4 mm RuPBA/Au region without ionogel coating. An acrylic adhesive backed plastic well was fabricated by laser cutting and attached to the device. The well was initially filled with 25 μL of 0.1 M sodium phosphate buffer, pH 5.5. A pipet was used to incrementally add 1 μL aliquots of the dopamine solution (0.1 mM dopamine hydrochloride, 0.1 M sodium phosphate buffer, pH 5.5) to the reservoir.

**X-ray diffraction**

Powder x-ray diffraction measurements were performed on a PANalytical Empyrean diffractometer equipped with a PIXcel3D detector and operated at 44 kV and 40 kA using Cu K$\alpha$ radiation ($\lambda$= 1.5406 Å). A reflection-transmission spinner was used as a sample holder and the spinning rate was set at 4 rpm. The patterns were collected in the 2$\theta$ range of 5 to 75°, and the step size was 0.026°.

**X-ray photoelectron spectroscopy**

Samples for x-ray photoelectron spectroscopy were prepared by dropcasting the RuPBA ink on a Au-coated $SiO_2$/Si substrate. The spectra were acquired using Al $K_\alpha$: 1486.6 eV source and monochromator (Scienta Omicron, Inc., XM 1000Mkii) under ultrahigh vacuum.

**RuPBA film conductivity and ECRAM pulsed-programming measurements**

Electrical measurements of Van der Pauw samples and pulsed conductance tuning of ECRAM devices were performed under $N_2$ atmosphere at slightly reduced pressure (~600 torr) and controlled temperature using a CPX-VF probe station and temperature controller (Lakeshore).

Sheet resistance measurements were performed with a Keysight B1500A semiconductor device analyzer to obtain conductivity based on the measured film thickness. Current-voltage recordings were also performed at each temperature to make sure the response remained ohmic.

Pulsed programming and state retention testing of ECRAM devices was performed using a custom LabVIEW waveform generation program, as previously described,[32,33,76] to control analog voltage outputs and record input current to an NI-DAQ PCIe-6363 at 10 kHz sampling. To switch the ECRAM device gate to open circuit during the 'read' steps, a CMOS analog switch (MAX327, MAXIM Integrated Products) was implemented and powered by a dual output DC power supply (E3620A, Keysight). A low-noise current amplifier (FEMTO, DLPCA-200) was used to measure the current at $10^6$ V/A gain and 200 kHz low-pass analog filter. The gate charge, $\Delta Q$, resulting from each write pulse was measured by integration of the raw current-time traces using the cumulative trapezoidal numerical integration function in MATLAB 2019.

**Cell Culture**

Neuro2a cells were purchased from ATCC (CCL-131) and cultured in high-glucose DMEM (Fisher Scientific #11995073) with 10% Fetal Bovine Serum (FBS) (Seradigm Premium Grade, Avantor) and 1% Penicillin/Streptomycin (Thermo Fisher #15140122). Cells were grown at 37 °C and 5% $CO_2$ and passaged/harvested using Trypsin-EDTA (Fisher Scientific #25300120).

**Growth and Viability Assays**

A 1 mg/mL aqueous dispersion of RuPBA was spotted onto wells of a 24-well plate (200 ul per well) and allowed to dry at room temperature for 1 hour. Cells were seeded at densities of 10,000, 25,000 and 50,000 cells/well in quadruplicate, and allowed to grow for 3 days, either in the presence or absence of RuPBA. At the end of 3 days, cells were harvested, stained with Trypan Blue (1:1 ratio, incubated for 5 minutes at room temperature) and counted using the BioRad TC20 Cell Counter. Pictures of the cells on the RuPBA substrate were taken using brightfield imaging with the EVOS M5000 imaging system (at 200X magnification).

**Scanning Electron Microscopy**

Cells were prepared for scanning electron microscopy as previously described.[77] Briefly, cells were plated on Nunc Thermanox coverslips (Thermo Fisher #150067) and grown for 3 days in DMEM media with 10% FBS (see above). The coverslips were then fixed in 2.5% glutaraldehyde (Fisher Scientific #O2957-1) at room temperature for 30 minutes, washed 3 x 2 minutes with rinsing buffer (phosphate buffer), and post-fixed with 1% osmium tetroxide (Sigma #20816-12-0). After another 2-minute wash with rinsing buffer and 2 x 2-minute washes with deionized water, the coverslips were dried with anhydrous ethanol. Samples were then dehydrated using a $CO_2$ critical point dryer, sputter coated with ~1 nm Au, and imaged using a FEI Nova NanoSEM 450 at 5 kV accelerating voltage.

**In-situ UV/Vis/NIR and Raman spectroscopy**

All spectroelectrochemical measurements were performed using a BioLogic SP-300 bipotentiostat.

In-situ UV/Vis/NIR spectroscopic interrogation of RuPBA films on ITO-coated glass slides (Sigma Aldrich) was performed using a JASCO V-770 spectrophotometer equipped with an integrating sphere. The spectroelectrochemical cell consisted of inkjet-printed RuPBA on ITO-coated glass (working electrode), a gold wire (counter electrode), a Ag/AgCl wire (quasireference electrode), and a quartz window in a custom-designed Teflon cell with rubber gaskets to seal electrolyte. The electrolyte contained 1.0 M LiFSI dissolved in $Pry_{14}TFSI$.

The electrochromic cell shown in the Video S1 of the Supplementary Information was constructed using two FTO-coated glass electrodes, each with inkjet-printed RuPBA films, and double-sided Kapton tape to seal the 1.0 M LiFSI in $Pry_{14}TFSI$ electrolyte between the two electrodes. The working electrode consisted of square shaped RuPBA printed film of 2.25 $cm^2$ area. The printed RuPBA area of the counter/reference electrode was a square frame with inner and outer square edge lengths of 3 and 4 cm (7 $cm^2$ frame area).

In-situ NIR absorption spectra of ECRAM devices were acquired with an NIRQuest+2.5 spectrometer from Ocean Insight. Fiber optic cables for 400-2400 nm wavelength range (Thorlabs, AFS50/125Y) were used with collimating lenses to direct the beam path of the tungsten-halogen light source through the ECRAM device using an adjustable stage for film transmission measurements (Stage-RTL-T, Ocean Insight). Blank reference measurements were performed by positioning the light beam (~0.2 mm cross-sectional diameter) at an area of the ECRAM device that included the glass microscope slide, PET substrate, and ionogel, but not RuPBA. The sample was then repositioned to pass the incident light through the middle of the RuPBA channel. Constant potentials were applied to the gate for 725 s using a BioLogic SP-300 bipotentiostat. Afterwards, the gate was set to open circuit and the channel conductance was monitored while simultaneously recording spectra (0.5 s integration time, 10 scans averaged for each spectrum).

In-situ Raman spectra were acquired with an inVia™ Raman microscope (Renishaw) using a 633 nm laser at 0.1% power. Using a 50X objective, the beam was focused to an area where the RuPBA channel film coats one of the Au contacts. The plasmonic enhancement from Au was found to be necessary to achieve sufficient signal. Acquisition settings: 50 s exposure time, 60 accumulations (3000 s total acquisition time per spectrum).

**DFT Computation**

All density functional theory calculations were performed using the Vienna ab initio simulation package (VASP).[78] The geometry and electronic properties were predicted using the hybrid HSE06 exchange-correlation functional within the spin polarization formulism. The planewave basis set energy cut-off was set to 500 eV and a 3x3x2 k-point grid was using during geometry optimization (atomic positions and cell volume). The geometries were optimized until all forces were below 0.02 eV/Å. A single-point calculation was then preformed, with a k-point grid of 6x6x4, to determine the density of states. The band structures and hole effective masses ($m_h$) were determined using AMSET[79] which depends on BoltzTrap2[80] for interpolation. Seekpath[81] was used to determine the high symmetry band paths; note, some paths have been omitted in Figures 6 and S9 for aesthetics reasons because they looked visually similar. The lowest $m_h$ determine via AMSET at gamma are reported. Three oxidation states were considered for both PB and RuPBA; fully reduced (*fr*), partially reduced (*pr*), and neutral. The different oxidation states were achieved by varying the number of Li atoms (4, 3 and 2 Li atoms respectively; note, PB calculations were performed with K in place of Li). An orthorhombic unit cell was used containing 4 Ru (see Fig. S7 of Supplementary Information) which is the smallest system allowing for a *pr* state to be represented.

The hole charge mobility was calculated for *pr*-RuPBA with the PBEsol+U functional using a Hubbard $U_{eff}$ value of 4.0 eV (resulting band gap is in good agreement with HSE06). The planewave basis set energy cut-off was set to 500 eV and a 4x4x3 k-point grid was used during geometry optimization; the geometry was relaxed until all forces were less 0.0005 eV/Å. The elastic constants, effective phonon frequency, static and high-frequency dielectric constants and deformation potential were calculated at the same level on theory. A signal-point calculation using a denser k-point mesh, 8x8x6, was used to obtain the wave function coefficients and for band structure interpolation. These properties/calculations were used to determine acoustic deformation, ionized impurity, and polar optical phonon scattering. The reported hole mobility is calculated for a carrier concentration of $1.8 \times 10^{21}$ cm$^{-3}$ which corresponds to 1 carrier per unit volume. The charge mobility calculation was performed using AMSET[79].

# Extended Data

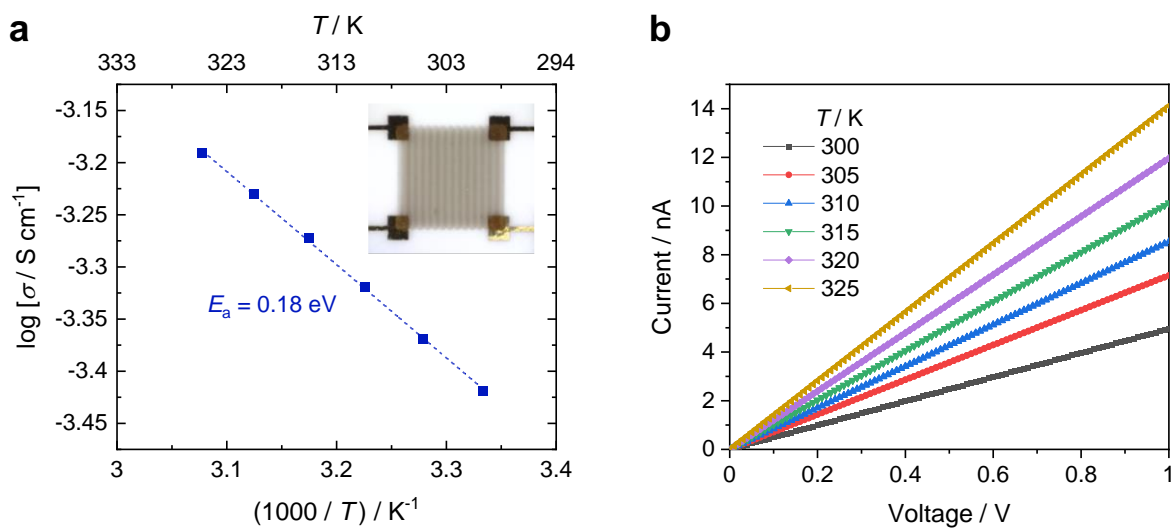

**Extended Data Fig. 1.** DC conductivity-temperature relationship for RuPBA at the as-synthesized valence state. **a**, Log(conductivity, $\sigma$) vs reciprocal temperature with corresponding activation energy ($E_a$) labelled on plot. Inset image shows the inkjet-printed 2×2 mm RuPBA film on a Au/PET van der Pauw square design, film thickness of 0.6 ±0.1 µm. **b**, Current-voltage recordings between two adjacent terminals of the van der Pauw sample at varied temperature.

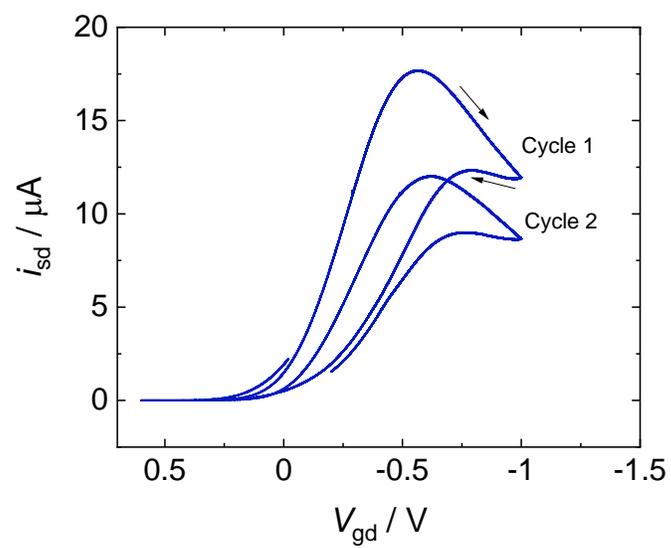

**Extended Data Fig. 2.** Transfer curve for a RuPBA/Li-ionogel device when the gate-drain voltage is scanned to -1 V at 30 µV/s, resulting in more partial oxidation of RuPBA in the channel in comparison to Fig. 1c. Source-drain bias of 0.1 V.

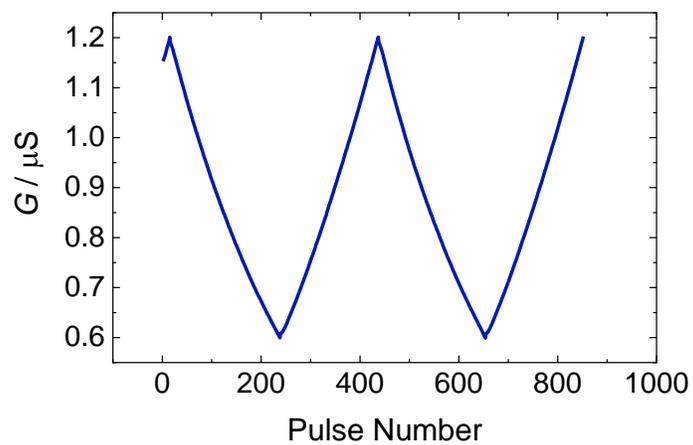

**Extended Data Fig. 3.** Channel conductance during pulsed programming at $V_{gd}$ = +1 V/-1 V for depression/potentiation at 10 ms gate pulse durations for an RuPBA device employing a proton-conducting ionogel electrolyte.

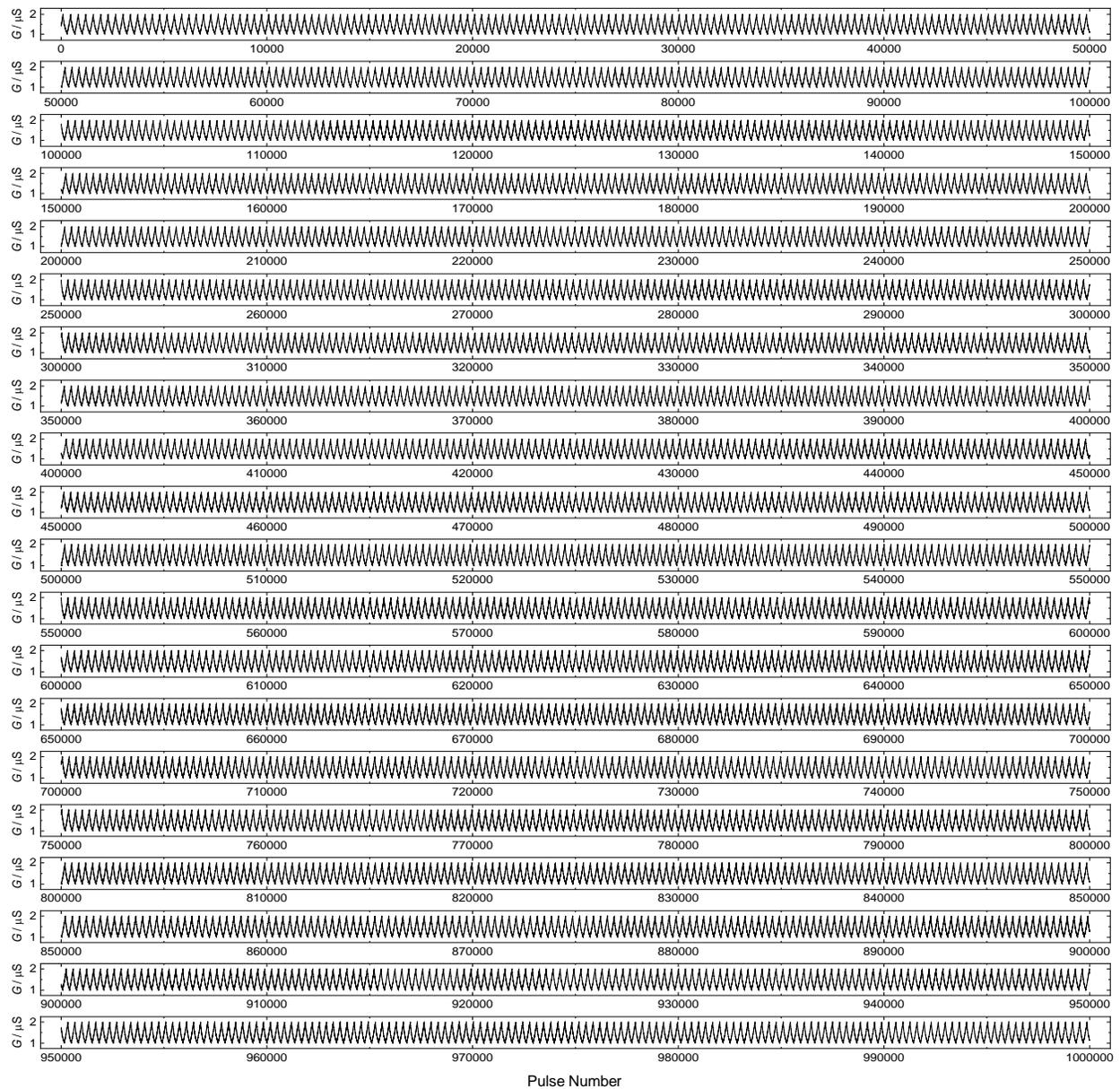

**Extended Data Fig. 4.** Full conductance trace resulting from one million write pulses and more than 6000 potentiation/depression ramps for a RuPBA EC-RAM device with Li ionogel electrolyte.

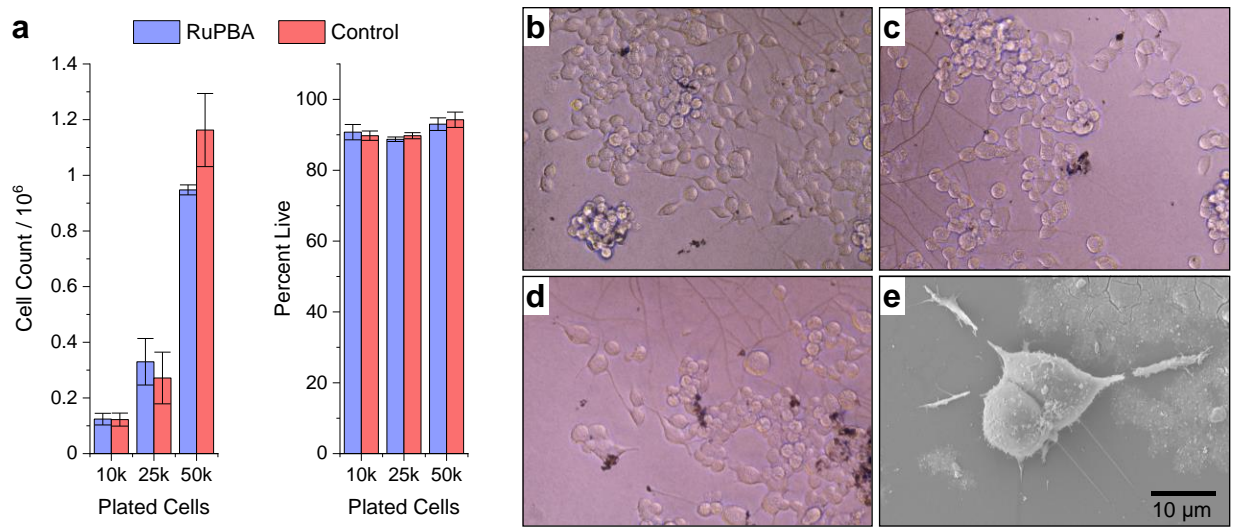

**Extended Data Fig. 5.** Compatibility of neuronal cells grown on RuPBA films. **a**, Growth and viability assays for Neuro2A cells grown on well plates modified with or without RuPBA films (blue or red columns, respectively). Left panel shows cell count after 3 days of incubation, each well initially seeded with 10,000, 25,000 or 50,000 cells, and right panel shows corresponding percentage of viable cells (percent live) based on Trypan Blue staining. Error bars represent the standard error of the mean based on 4 replicates. **b-d**, Optical micrographs and **e**, SEM image of Neuro2A cells grown on RuPBA films.

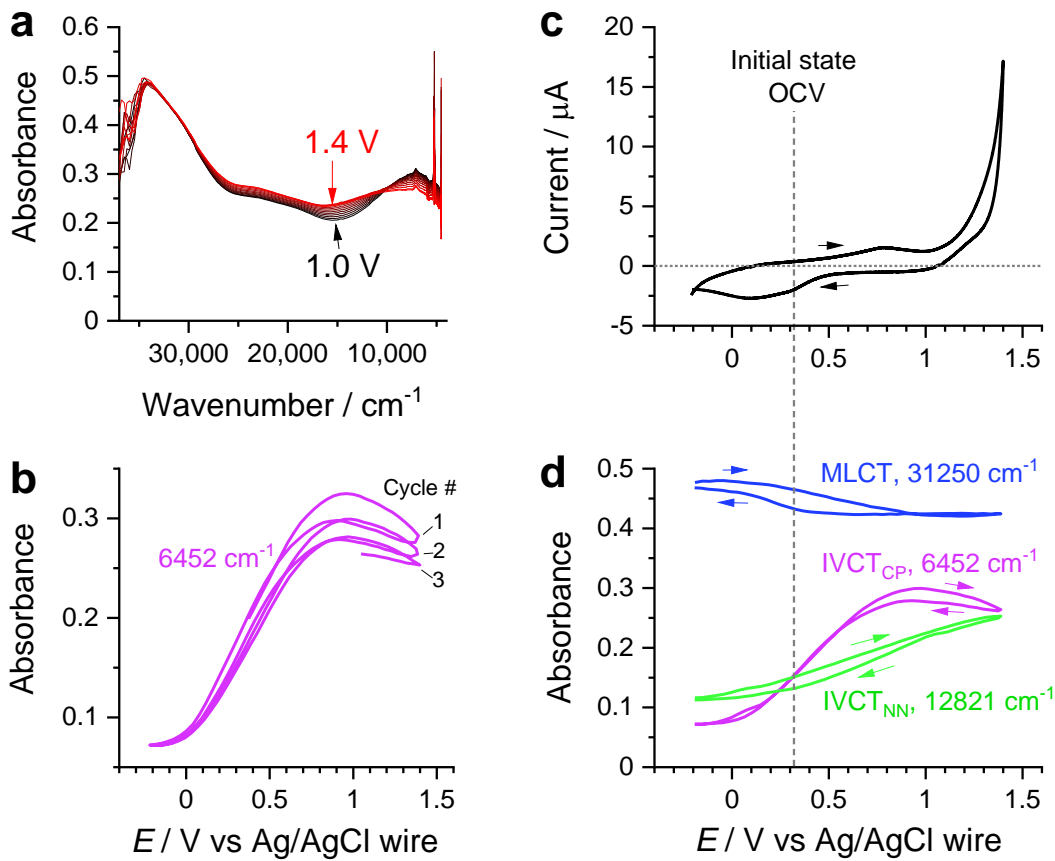

**Extended Data Fig. 6.** UV/Vis/NIR spectroscopic monitoring of redox processes for a RuPBA film on ITO-coated glass electrode in a cell containing 1.0 M LiFSI dissolved in Pyr$_{14}$TFSI ionic liquid. **a**, Overlaid absorption spectra recorded while scanning the potential from 1.0 to 1.4 V. Current-potential response from cyclic voltammetry at 50 µV/s, starting at -0.2 V (second cycle). Vertical dashed line marks the OCV of the initial state (315 mV) prior to performing the voltammetric scans. Overlaid absorption spectra recorded while scanning the potential from **b**, -0.2 to 1.0 V, and **c**, 1.0 to 1.4 V. **d**, Absorbance vs potential as recorded at three different wavelengths during the scan. Potential is reported versus Ag/AgCl wire reference electrode.

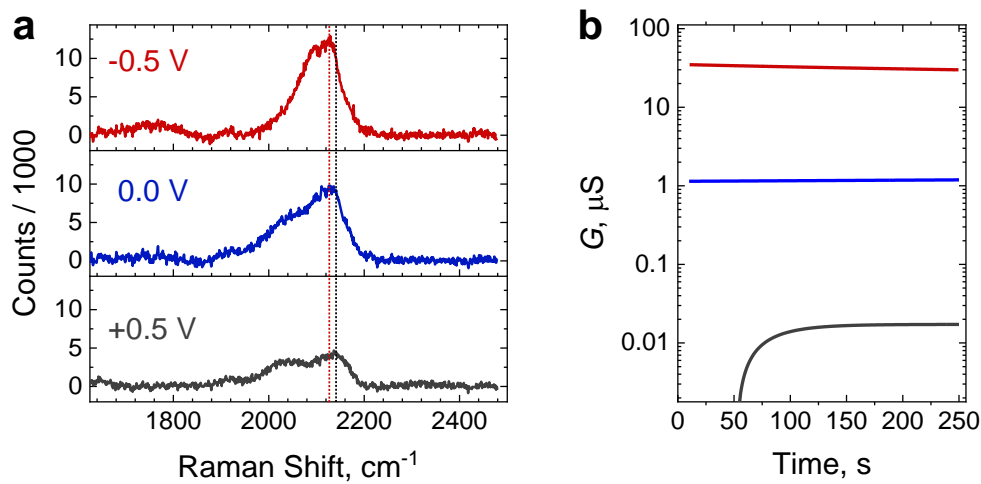

**Extended Data Fig. 7.** In-situ Raman spectroscopy of ECRAM device channel. **a,** Raman spectra of cyanide stretch region recorded in the device channel after charging for 30 minutes at 3 different gate voltages, as labeled on the plot. Vertical lines drawn as a guide for the peak maxima for the most oxidized channel at $V_{gd}$ = -0.5 V (red dotted line) and the most reduced channel at $V_{gd}$ = +0.5 V (blue dotted line). **b,** Conductance traces (log scale) recorded during Raman scan after gate charging.

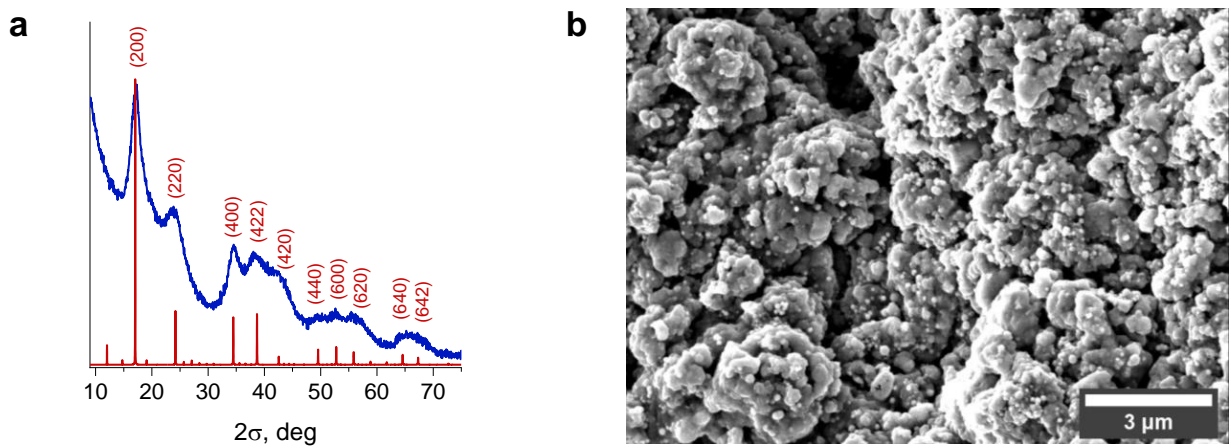

**Extended Data Fig. 8.** Structure and morphology of RuPBA. **a**, Observed powder x-ray diffraction pattern (blue) and calculated[82] pattern (red), assuming face-centered cubic symmetry and a lattice parameter of 10.4 Å, with Miller indices labeled on the plot for the dominant lattice planes. **b**, Scanning electron micrograph of RuPBA.

| $E$ / V | $\lambda$ / eV | $h\Delta\nu_{1/2}$ /eV | $H_{\text{eff}}$ / eV | $E_a^*$ / meV | $\nu_{\text{el}}$ / $10^{14}$s$^{-1}$ | $k_{\text{ct}}$ / $10^{12}$s$^{-1}$ | $\mu_h$ / cm$^2$V$^{-1}$s$^{-1}$ |
|---|---|---|---|---|---|---|---|
| 0.00 | 0.81 | 0.64 | 0.20 | 50 | 7.6 | 8.9 | 0.95 |
| 0.05 | 0.76 | 0.50 | 0.25 | 24 | 12 | 25 | 2.7 |
| 0.10 | 0.74 | 0.45 | 0.26 | 16 | 13 | 34 | 3.7 |
| 0.15 | 0.66 | 0.39 | 0.24 | 11 | 12 | 41 | 4.3 |
| 0.30 | 0.59 | 0.36 | 0.22 | 11 | 10 | 42 | 4.5 |
| 0.60 | 0.51 | 0.34 | 0.18 | 11 | 7.5 | 41 | 4.3 |
| 1.00 | 0.55 | 0.39 | 0.18 | 17 | 7.3 | 33 | 3.5 |

**Extended Data Table 1.** Measured parameters from deconvolution of IVCT$_{\text{CP}}$ bands at varied applied potentials, as determined from the spectra shown in Fig. 2a. Parameters are defined in the main text and Section S6 of the Supplementary Information.

| Oxidation State | Lattice Constants (Å) | | | Band Gap (eV) | | $m_h$ |
| --- | --- | --- | --- | --- | --- | --- |
| | a | b | c | direct | indirect | |
| neutral | 7.269 | 7.293 | 10.334 | 1.26 | 0.98 | 0.70 |
| partially reduced | 7.415 | 7.217 | 10.295 | 0.50 | 0.10 | 0.41 |
| fully reduced | 7.543 | 7.038 | 10.180 | 4.29 | 4.29 | 0.52 |

**Extended Data Table 2.** DFT/HSE06 optimized lattice parameters, direct and indirect band gaps, and hole effective mass for RuPBA in three different oxidation states.

## Acknowledgements

This research was supported by the Laboratory-Directed Research and Development (LDRD) Program. AAT and MJM were also supported by the DOE Office of Science Research Program for Microelectronics Codesign (sponsored by ASCR, BES, HEP, NP, and FES) through the Abisko Project, PM Robinson Pino (ASCR). Sandia National Laboratories is a multimission laboratory managed and operated by National Technology and Engineering Solutions of Sandia, LLC., a wholly owned subsidiary of Honeywell International, Inc., for the U.S. Department of Energy's National Nuclear Security Administration under Contract DE-NA-0003525. The views expressed in this article do not necessarily represent the views of the U.S. Department of Energy or the United States Government.

# Supplementary Information

# Tunable intervalence charge transfer in ruthenium Prussian blue analogue enables stable and efficient biocompatible artificial synapses

Donald A. Robinson, Michael E. Foster, Christopher H. Bennett, Austin Bhandarkar, Elizabeth R. Webster, Aleyna Celebi, Nisa Celebi, Elliot J. Fuller, Vitalie Stavila, Catalin D. Spataru, David S. Ashby, Matthew J. Marinella, Raga Krishnakumar, Mark D. Allendorf, and A. Alec Talin[*]

*Corresponding author: aatalin@sandia.gov

## Contents

## S1. Charge analysis of RuPBA ECRAM transfer characteristics

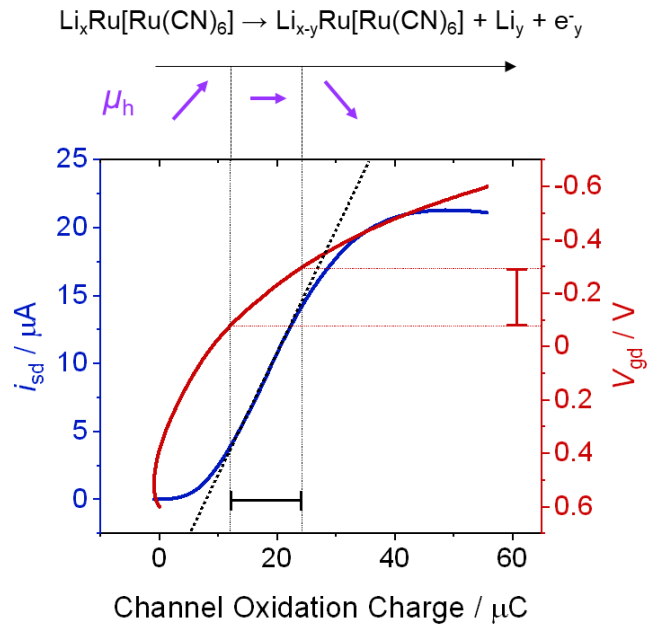

**Fig. S1.** Relationship between source-drain current and electrochemical charge (channel oxidation charge) during the oxidation of the RuPBA channel from $V_{gd} = 0.6$ V to -0.6 V as shown in Fig. 1c of the main text. The range of RuPBA oxidation states where $i_{sd}$ increases linearly with accumulated charge is encompassed within the vertical dashed lines and the corresponding $V_{gd}$ range is highlighted by the horizontal red dashed lines. The purple arrows indicate gate voltage regions when the mobility ($\mu_h$) increases, remains constant, or decreases during RuPBA oxidation.

## S2.   ECRAM device architecture and fabrication

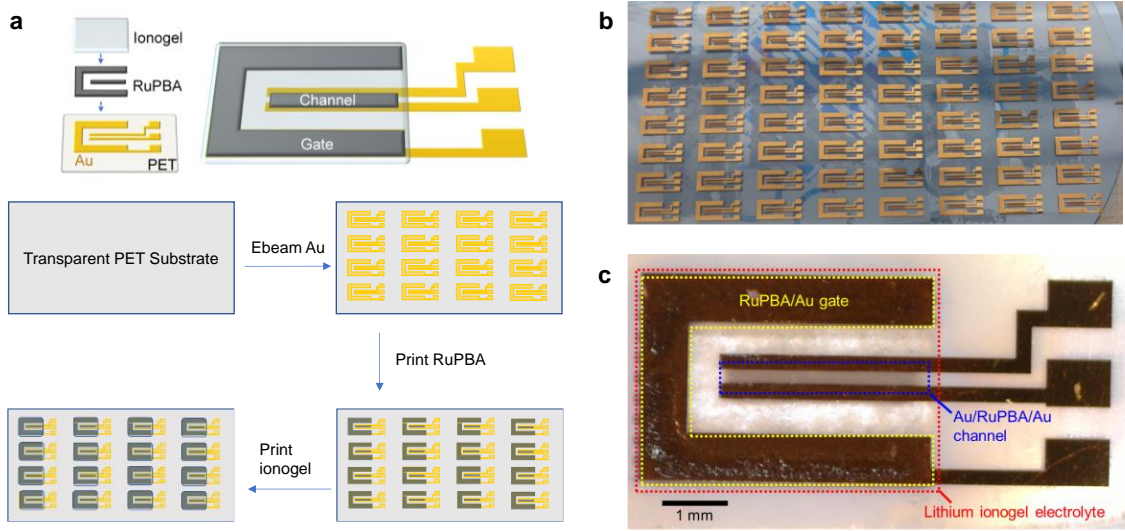

**Fig. S2.**  ECRAM device fabrication by inkjet printing. **a**, Device architecture and flow chart of fabrication process. **b**, Photo of device array on transparent PET substrate. **c**, Photo of individual printed device.

## S3. Simulation of ECRAM synapse array for online learning accuracy predictions

We evaluated the implications for online learning by simulating a hardware multi-layer perceptron, physically constructed of crossbars of various sizes using the CrossSim codebase (Table S1) populated with simulated RuBPA synapses learning with stochastic gradient descent (batch size 1). Synaptic switching linearity and symmetry was based on look up tables (Fig. S3) generated from the experimental data (Extended Data Fig. 4).

**Table S1. Neural network hyper-parameters used for training**

| Task | Topology (Input x Hidden Layer x Output) | Crossbar Core Type | Learning Rate (L.R.) | Other comments |
|---|---|---|---|---|
| Iris | (4 + bias) x 8 x 4 | Balanced (2 Devices per Syn.) | $\alpha$=5e-3 | Constant L.R. |
| OCR | (64 + bias) x 36 x 10 | Balanced | $\alpha$=8e-3 | Constant L.R. |
| MNIST | (784 + bias) x 300 x 10 | Balanced | $\alpha$=2.5e-3 | L.R. schedule ($\alpha$/2 epochs 3-5, $\alpha$/3 epochs 6-8, $\alpha$/4 epochs 9-11, $\alpha$/5 after) |
| f-MNIST | (784 + bias) x 300 x 10 | Balanced | $\alpha$=1e-3 | L.R. schedule (same as MNIST) |

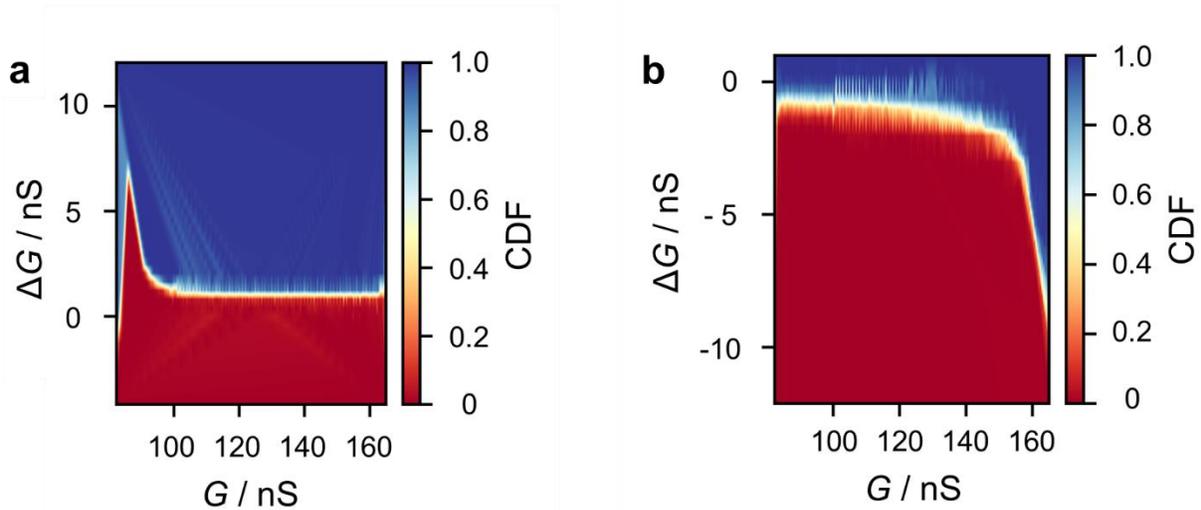

**Fig. S3.** Maps of RuPBA ECRAM switching magnitude and probability as a function of conductance (look up tables) during **a**, potentiation and **b**, depression, originating from recording shown in Extended Data Fig. 4. Median probability is the yellow/white region along the middle.

Our results are comparable to that of other leading linear and symmetric nonvolatile memory (NVM) candidates, as show comprehensively in Table S2. RuPBA devices achieves parity with or slightly outperforms synapses based on PEDOT/PSS,[1,2] lithium cobalt oxide (LCO),[3] and bulk-RRAM devices based on $TiO_{2-x}$.[4] While our simulations do not incorporate realistic device-to-device effects as in Li *et al.*,[2] additional agreement with numeric training can be achieved by centering synapses to a tighter, shared workable range. As this range is the most linear and symmetric range regardless of RuBPA geometry, this augurs well for the future of these devices in online learning and inference contexts. Along with a variety of promising NVM devices with a degree of high linearity and symmetry,[4,5] these results establish RuBPA ECRAM devices as a leading candidate for future embedded/edge systems that require online learning capabilities.

**Table S2. Comparison to other NVM device results**

| Device | Small Digits (OCR) Task | Large Digits (MNIST) Task | Fashion MNIST task |
|---|---|---|---|
| *Numeric* | *96.2%* | *98.1%* | *79.4%* |
| RuBPA- (this work) | 94.9% | 97.3% | 78.9% |
| PEDOT:PSS- Original[1] | 95.2% | 96.5% | -- |
| PEDOT:PSS- Single[2] | 95% | 95.9% | 79.1% |
| PEDOT:PSS- Multiple with Centering[2] | 93.8% | 92.9% | 70.3% |
| LCO[3] | 96% | 97.5% | -- |
| Bulk-RRAM ($TiO_{2-x}$, YSZ active layer)[4] | -- | 97.5% | -- |
| silicon-oxygen-nitrogen-oxygen-silicon (SONOS)[6] | -- | 95.8% | -- |

## S4. Analysis of switching time and energy for ECRAM synapse scaling relationship

To evaluate the upper limit to the device programming rate, one must first determine a threshold criterion for the minimum conductance change, $\Delta G_{min}$, that can be accurately measured. The number of discrete programmable conductance states within a specified range ultimately depends on the accuracy of the measured conductance, which is dictated by the total noise level of the read current. The standard deviation from all measured $\Delta G$ values for multiple potentiation-depression pulsed programming cycles ($\sigma_{\Delta G}$) was selected to set the threshold criterion as $\Delta G \geq 2\sigma_{\Delta G}$ with the limiting case, $\Delta G_{min} = 2\sigma_{\Delta G}$. For a fair comparison, the applied write pulse durations ($\Delta t_w$) were varied for each device size to achieve ~70 pulses per ramp, giving roughly the same average change in conductance per switching pulse ($\Delta G_{avg}$). The average $\sigma_{\Delta G}$ for the three different device sizes was 0.5(±0.1) nS, giving $\Delta G_{min} \approx 1$ nS (See Table S1). To evaluate switching speed, we estimate the minimum write pulse duration, $\Delta t_{min}$, based on the average programming rate, $\Delta G_{avg}/\Delta t_w$, to give $\Delta G_{avg}/\Delta t_w = \Delta G_{min}/\Delta t_{min}$, which leads to

$$\Delta t_{min} = \frac{\Delta G_{min}}{\Delta G_{avg}} \times \Delta t_w \quad . \tag{S1}$$

The minimum energy requirement for each weight update, $\Delta E_{min}$, is the product of the voltage applied to the gate, $V_g = 1$ V, and the amount of transferred charged, $\Delta Q_{min}$, to achieve the minimum conductance change ($\Delta E_{min} = V_g \Delta Q_{min}$). To determine $\Delta Q_{min}$, we first measured each increment of transferred charge, $\Delta Q$, by integration of the current-time response for each write pulse (eq 3).

$$\Delta Q = \int_{t_1}^{t_2} i_w \, dt \tag{S2}$$

We then use the $\Delta G_{min}/\Delta G_{avg}$ ratio to approximate $\Delta Q_{min}$ from the measured average charge, $\Delta Q_{avg}$.

$$\Delta Q_{min} \cong \frac{\Delta G_{min}}{\Delta G_{avg}} \times \Delta Q_{avg} \tag{S3}$$

**Table S3. Measured Parameters Used to Determine Minimum Switching Time and Energy**

| $L$ / μm | $\Delta t_w$ / ms | Total pulses | States per ramp | $\Delta G_{avg}$ / nS | $\sigma_{\Delta G}$ / nS | $\Delta G_{min}$ / nS | $\Delta t_{min}$ / ms | $\Delta Q_{avg}$ / nC | $\Delta Q_{min}$ / nC | $\Delta E_{min}$ / nJ |
|---|---|---|---|---|---|---|---|---|---|---|
| 100 | 0.75 | 23569 | 83 | 1.0 | 0.5 | 1 | 0.7 ±0.4 | 1.51 ±0.02 | 0.8 ±0.4 | 0.8 ±0.4 |
| 300 | 6.00 | 9528 | 64 | 1.3 | 0.4 | 1 | 5 ±1 | 16 ±1 | 6 ±2 | 6 ±2 |
| 400 | 10.00 | 8265 | 73 | 1.1 | 0.6 | 1 | 9 ±5 | 28 ±1 | 13 ±6 | 13 ±6 |
| | | Mean | 73 | 1.1 | 0.5 | | | | | |
| | | Std. dev. | 9 | 0.1 | 0.1 | | | | | |

## S5. Detailed description of dopamine sensor operation

Fig. S4a shows the simultaneously recorded gate voltage and current that accompanies the source-drain conductance trace in Fig. 3d of the main text for dopamine sensing measurements. During the first 590 s, the device reservoir contains phosphate buffer without dopamine. The first applied gate voltage pulse ($p_1$) switches the conductance to a lower value and the second pulse ($p_2$) returns the conductance to the initial value ($G_0 = 4.85$ μS). The open-circuit voltage between the gate and channel ($V_g$) responds similarly, switching to a more positive value after $p_1$ and returning to the initial value (-80 mV) after $p_2$. Upon introduction of 4 μM DA (step 1), $V_g$ shifts to more negative potential as the dopamine undergoes redox with the RuPBA at the gate because the decrease of mixed valence II:III sites causes the Fermi level of the gate to increase (Fig. S4b). Meanwhile, the channel conductance remains at 4.85 μS because the gate is still at open circuit. When the next $p_1$ pulse is applied (step 2), the gate oxidation current jumps to a higher magnitude than in the previous three $p_1$ pulses because the dopamine reaction generated $Ru^{II}$ sites to (re)oxidize, thus lowering the Fermi level of the gate and causing the opposite reaction (reduction) at the channel. After the application of $p_2$, the gate-channel circuit is switched open (step 3), and the resulting conductance is lower than the previous value when no dopamine was present. The net result of steps 1-3 is to increase $Ru^{II}$ sites and decrease conductivity at all regions of the device, starting with the gate (step 1) and ending with the channel (steps 2-3).

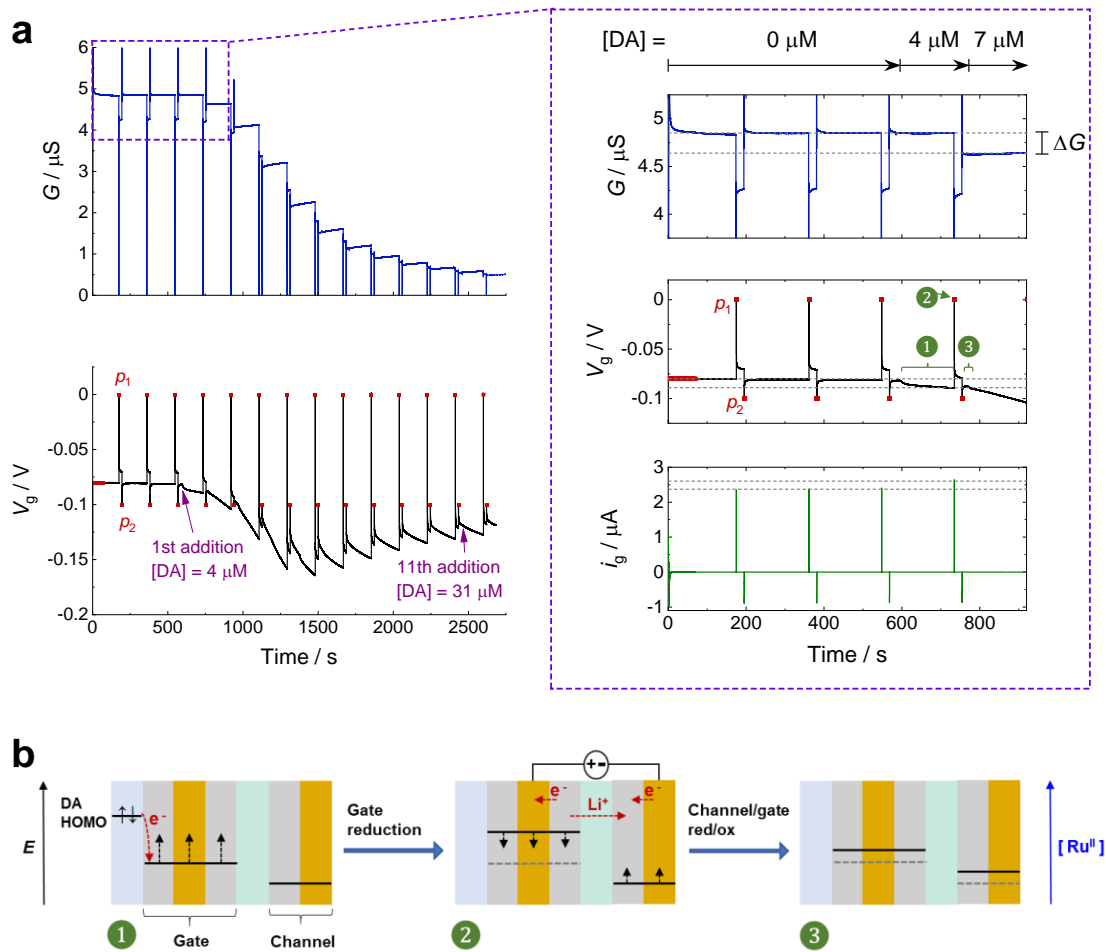

**Fig. S4.** RuPBA ECRAM dopamine sensor operation. **a**, Recordings of channel conductance ($G$), gate voltage ($V_g$), and gate current ($i_g$) during incremental additions of DA. Black sections of the $V_g$ trace show when the voltage was measured at open-circuit and red points mark when short voltage pulses were applied to the gate ($p_1$: 0.0 V, 100 ms, $p_2$: -0.1 V, 435 ms). **b**, Schematic depicting how the electrochemical potential and Ru$^{II}$ concentration change at the gate and channel at different stages of operation. The steps in part **b** are labeled in the zoomed-in section of the $V_g$ trace of part **a** for the first addition of DA solution. $V_{sd}$ = 0.1 V.

## S6. IVCT spectral analysis based on electron transfer theory

Spectroscopic determination of the reorganization energy, electronic coupling, and activation energy followed the methods detailed by Brunschwig, Creutz, and Sutin based on a two-state model for a class 2B system.[7] The IVCT band manifold ($v_{max} = E_{op}/h$) represents the combination of reorganization energy ($\lambda$) and the energy difference between the reactant and product ground states ($\Delta E_0$). For IVCT$_{CP}$, $\Delta E_0 = 0$ because the reactant and product states are essentially identical, leading to eq S5 to determine $\lambda$.

$$hv_{max} = E_{op} \approx \lambda \tag{S5}$$

For such symmetrical systems, one can approximate the electronic coupling ($H_{eff}$) based on the peak shape following a two-state model (eq S6)

$$\Delta v_{1/2} = [4\ln(2)\lambda k_B T]^{1/2} + \lambda - 2H_{eff} \tag{S6}$$

where $\Delta v_{1/2}$ is the full width at half maximum for the fitted IVCT$_{CP}$ band. The activation energy, $E_a^*$, was calculated by eq S7.

$$E_a^* = \frac{(\lambda - 2H_{eff})^2}{4\lambda} \tag{S7}$$

This equation modifies the more well-known expression for self-exchange reactions[8] ($E_a = \lambda/4$) by including the influence of electronic coupling on the activation energy. For a Robin-Day class 2 system, $\lambda > 2H_{eff}$ and $E_a^* > 0$. For class 3, $\lambda < 2H_{eff}$ and the activation barrier does not exist.

The electronic transmission coefficient, $\kappa$, describes the probability of electron transfer when the energy barrier is minimized according to the Landau-Zener model,[8,9] as expressed by eq S8,

$$\kappa = \frac{2 - 2\exp\left(-\frac{v_{el}}{2v_n}\right)}{2 - \exp\left(-\frac{v_{el}}{2v_n}\right)} \tag{S8}$$

where $v_{el}$ is the electronic frequency factor, calculated by eq S9.[8]

$$v_{el} = \frac{2\pi H_{eff}^2}{\hbar\sqrt{4\pi\lambda k_B T}} \tag{S9}$$

In spectroelectrochemical measurements (Fig. 4a of the main text), the spectrum of fully reduced RuPBA was used as a reference to subtract from spectra of the partially oxidized states, giving Fig. S5a. This was done to isolate the IVCT bands from the transitions that do not change during oxidation. This corrected absorbance was divided by wavenumber (Abs/$v$) to give the 'reduced' spectra,[10] as shown in Fig. S5c, which provides a more accurate determination of $\lambda$ and $H_{eff}$,[7] and the bands were fit as asymmetric gaussian peaks. Example deconvolutions and graphically represented measurements of the IVCT$_{CP}$ band parameters are shown in Fig. S6. The IVCT$_{CP}$ peaks for the reduced spectra were more ambiguous at higher potentials because $v_{max}$ shifts to a lower energy than the experimental cutoff (Fig S6b). To best approximate the partial peak width at the lower energy side, the fitted peak was constrained such that Abs/$v$ approaches zero at $v = 0$. This approach was guided by the observation of decreasing symmetry and broadness of the unfitted spectra at the lower potential range (ca. -0.1 to 0.15 V, as shown in the inset of Fig. S5c). Similar to the IVCT$_{CP}$ fit, the best fit for IVCT$_{NN}$ became more asymmetric at more oxidizing potentials as the coupling increases. The measured values of $\lambda$ and $\Delta v_{1/2}$ were used to determine $H_{eff}$ and $E_a^*$, as listed in Extended Data Table 1. A similar method was followed for the spectrum of as-synthesized RuPBA, as shown in Fig. 5a of the main text.

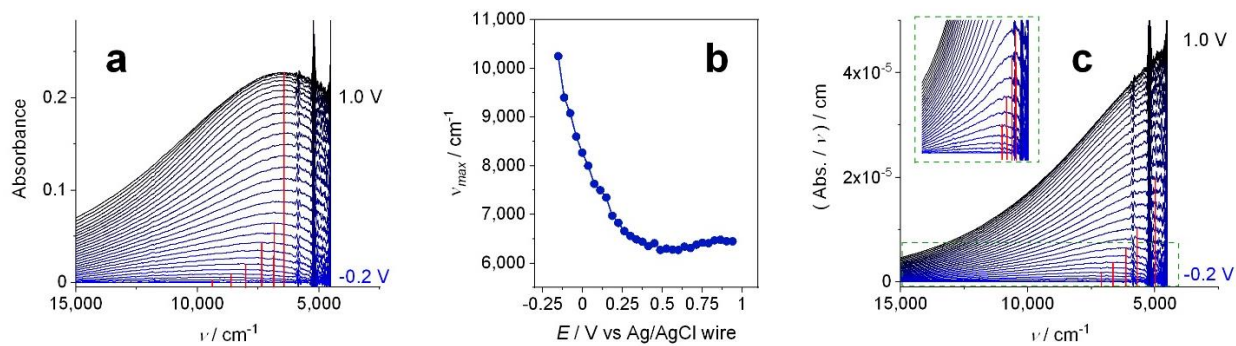

**Fig. S5.** Detailed analysis of absorption spectrum evolution during RuPBA oxidation. **a**, Spectra from Fig. 2a in the main text after subtracting the spectrum of *fr*-RuPBA at -0.2 V. Vertical red lines mark the absorbance maxima at different oxidation states. **b**, Measured wavenumber for the band manifold as a function of applied potential. **c**, 'Reduced' spectra from part **a** above after dividing by $v$.

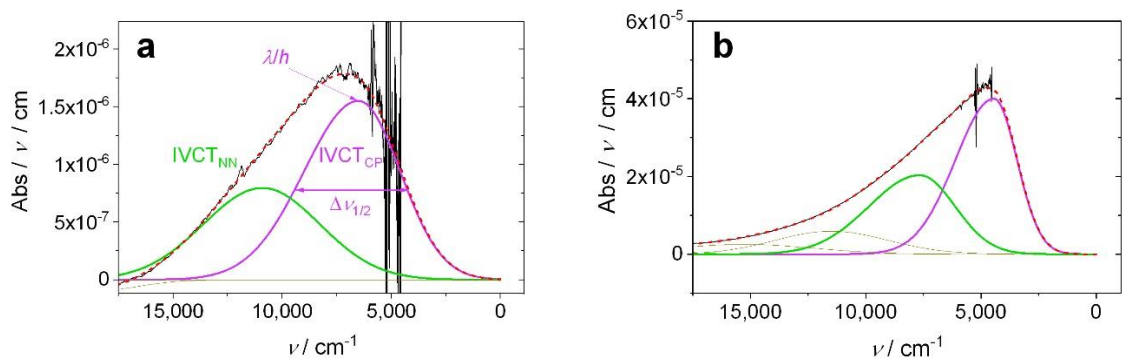

**Fig. S6.** Deconvolution of IVCT bands for reduced absorption spectra collected at **a**, $E = 0$ V and **b**, $E = 1.0$ V. These examples were taken from the overlaid spectra shown in Fig. 2a of the main text after subtracting the absorbance spectrum at $E = -0.2$ V and dividing the absorbance by wavenumber.

## S7. Predictions from DFT

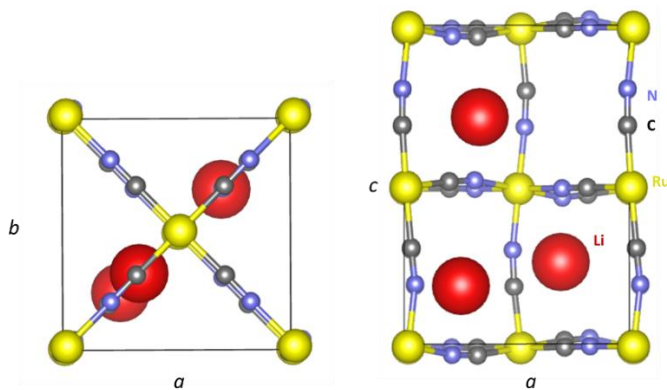

**Fig. S7.** DFT/HSE06 optimized geometry of partially reduced RuPBA, $Li_3Ru_2[Ru(CN)_6]_2$. Lattice parameters are reported in Extended Data Table 2.

**Table S4.** DFT/HSE06 optimized lattice parameters, direct and indirect band gaps, and hole effective mass for PB in three different oxidation states

| Oxidation State | Lattice Constants (Å) | | | Band Gap (eV) | | $m_h$ |
|---|---|---|---|---|---|---|
| | a | b | c | direct | indirect | |
| neutral | 7.081 | 7.317 | 10.182 | 2.29 | 2.19 | 1.3 |
| partially reduced | 7.237 | 7.285 | 10.267 | 0.36 | 0.28 | 2.2 |
| fully reduced | 7.288 | 7.322 | 10.341 | 3.15 | 3.15 | 2.9 |

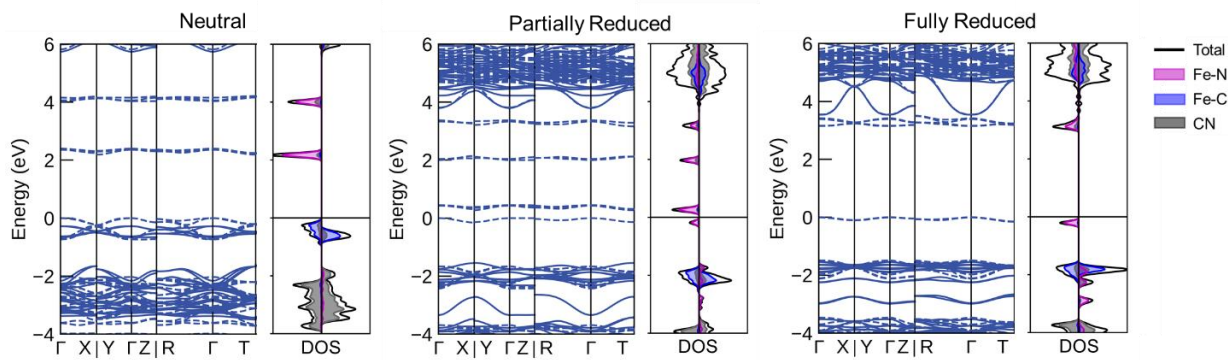

**Fig. S8.** DFT/HSE06 band structures and density of states of PB in the neutral, partially reduced and fully reduced states. The hole effective masses are reported in Table S4.

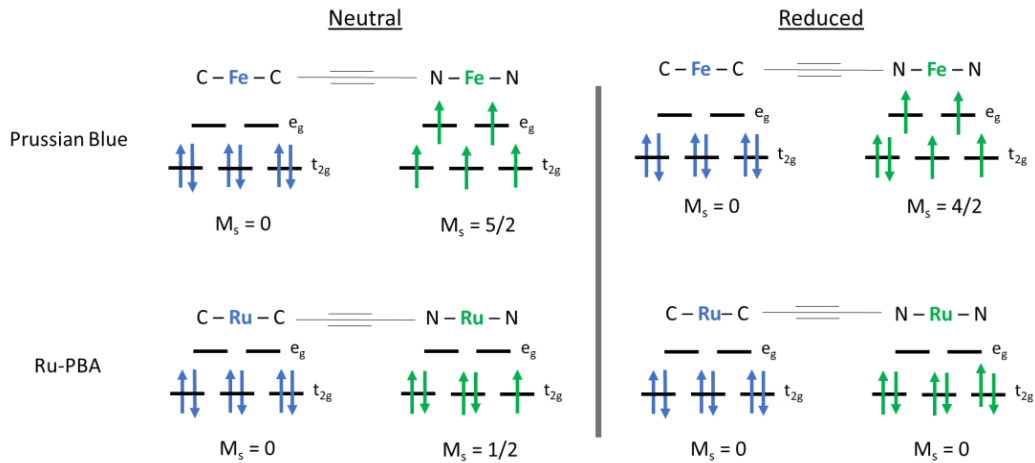

**Fig. S9.** Schematic of the electronic configuration of PB and Ru-PBA in the neutral and fully reduced states.

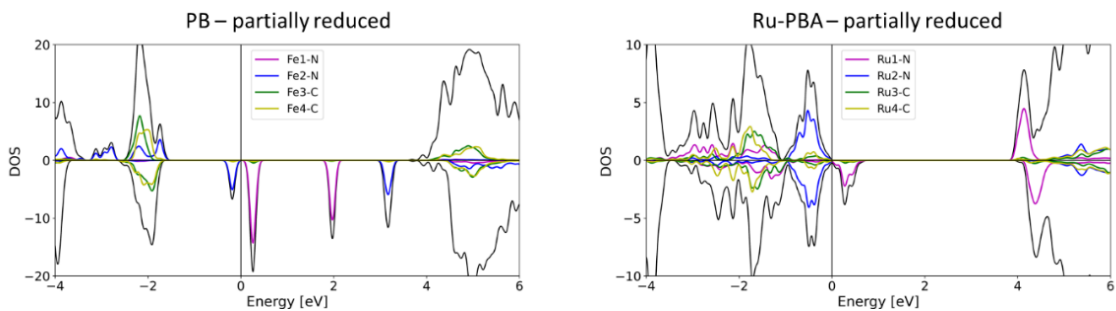

**Fig. S10.** DFT/HSE06 atom projected DOS for PB and Ru-PBA in the partially reduced state; the atom projections are shown for the 4 metal atoms in the calculation.

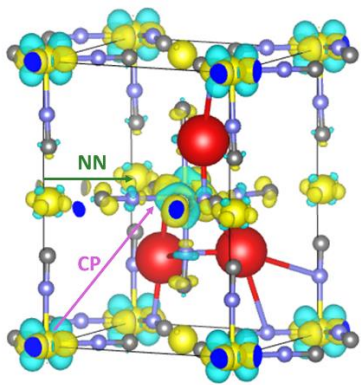

**Fig. S11.** DFT/HSE06 charge density difference for adding a hole (removing an electron) to *pr*-RuPBA. The figure illustrates that the hole delocalizes; moreover, the majority of the density is on the two RuN sites in the cell but some is on a RuC site. The delocalization, in the NN direction, is believed to be why the effective mass is lower for RuPBA than PB.

## S8. Optical profilometry and roughness mapping of inkjet-printed RuPBA film

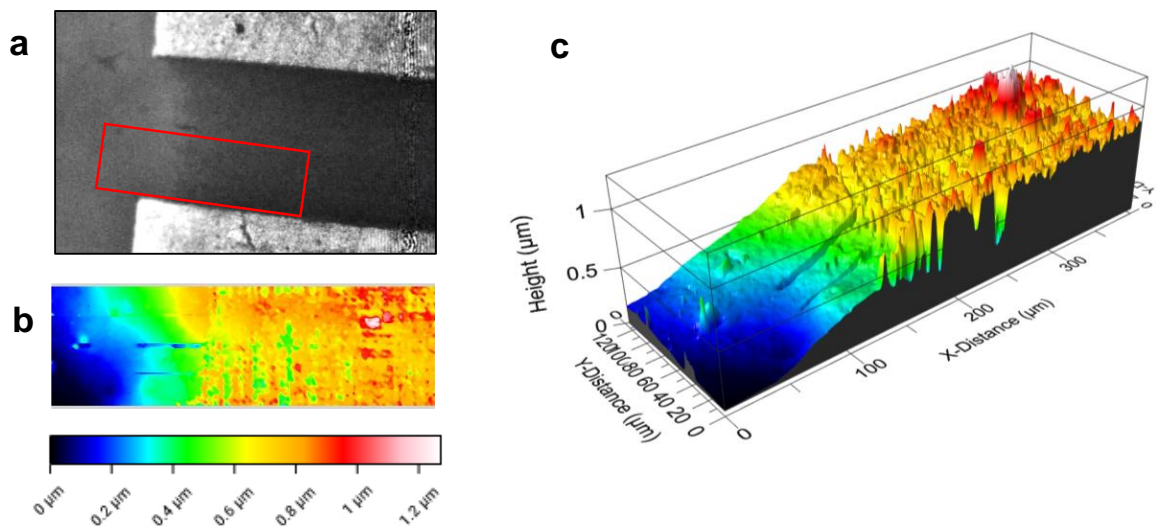

**Fig. S12.** Optical profiles of RuPBA channel for an EC-RAM device without electrolyte. **a**, Optical image showing selected region of the channel to map. **b,** Heat map and **c**, corresponding 3-D profile. Blue region indicates the PET substrate. Measured film thickness is $0.6 \pm 0.1$ µm based on 6 measurements at different edge regions of the channel film.

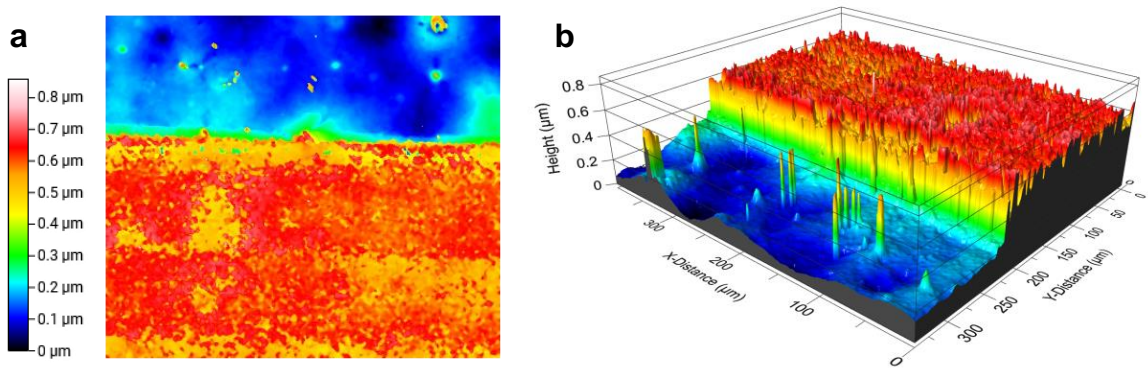

**Fig. S13.** Optical profiles of RuPBA Van der Pauw film for conductivity measurements. **a**, Example of heat map and **b,** corresponding 3-D profile focused at a section showing the edge of the film. Blue region indicates the PET substrate. Measured film thickness is $0.6 \pm 0.1$ µm from 15 measurements at different edge regions of the channel film.

## S9. X-ray photoelectron spectroscopy

Figure S13 shows the photoelectron spectra for as-synthesized RuPBA. The maximum intensity for Ru $3d_{5/2}$ occurs at a binding energy (BE) of 280.5 eV, which is close to that reported for $K_4Ru^{II}(CN)_6$ (280.9 eV),[11] suggesting a mostly reduced state, consistent with the high potassium content from elemental analysis (see Methods) and spectroscopic observations discussed in the main text. The Ru regions were best fit with two peaks of BE separated by ~1 eV (red and blue peak fits). The components at higher BE (blue fits) may at least partially correspond to $Ru^{III}$ content, but the relative intensity seems too high to assign to charge-localized $Ru^{III}$ sites considering the high degree of electronic delocalization in RuPBA. Also, the BE difference is half the expected value of 2 eV for comparing $Ru^{II}$ to $Ru^{III}$ in complexes with identical composition.[12] We therefore tentatively assign the two components to different coordination with either the N or C atoms of the cyanide bridging ligand, in addition to $H_2O$ bound at defect sites.

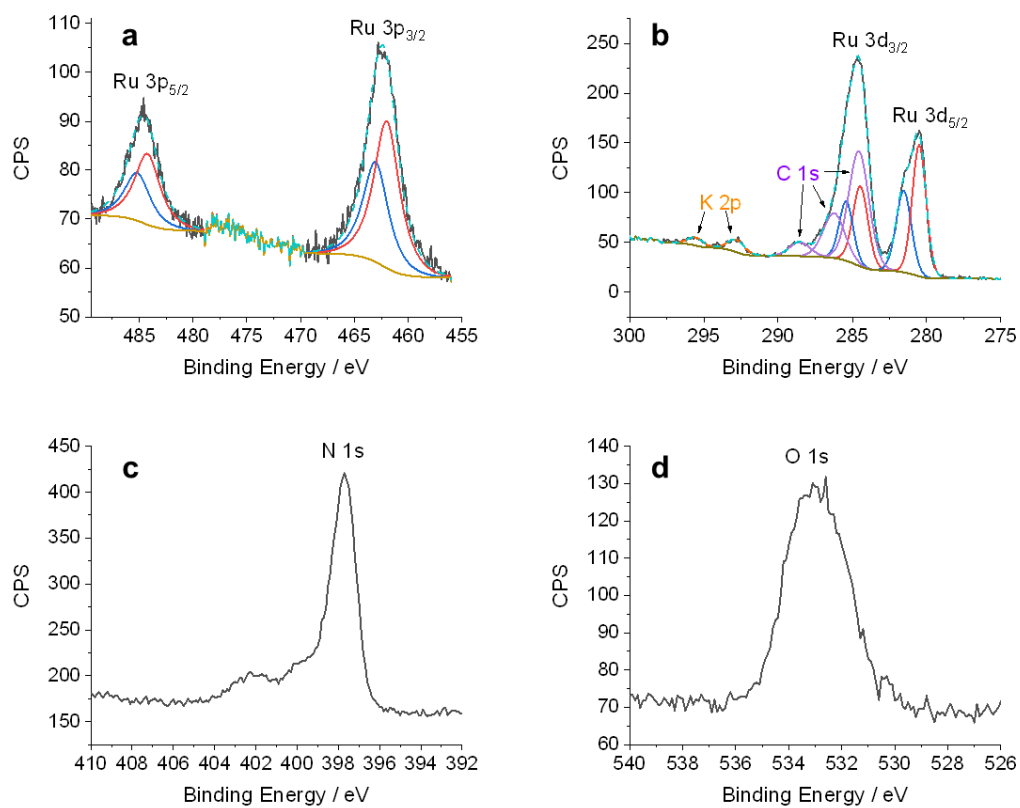

**Fig. S14.** X-ray photoelectron spectra of as-synthesized RuPBA showing **a**, Ru 3p, **b**, Ru 3d, **c**, N 1s, and **d**, O 1s regions.